\documentclass{jfm}
\usepackage{natbib,amsmath,amssymb,graphicx}
\usepackage[usenames,dvipsnames,svgnames,table]{xcolor}
\usepackage{braket}

\ifCUPmtlplainloaded \else
  \checkfont{eurm10}
  \iffontfound
    \IfFileExists{upmath.sty}
      {\typeout{^^JFound AMS Euler Roman fonts on the system,
                   using the 'upmath' package.^^J}%
       \usepackage{upmath}}
      {\typeout{^^JFound AMS Euler Roman fonts on the system, but you
                   dont seem to have the}%
       \typeout{'upmath' package installed. JFM.cls can take advantage
                 of these fonts,^^Jif you use 'upmath' package.^^J}%
      }
  \else
  \fi
\fi

\ifCUPmtlplainloaded \else
  \checkfont{msam10}
  \iffontfound
    \IfFileExists{amssymb.sty}
      {\typeout{^^JFound AMS Symbol fonts on the system, using the
                'amssymb' package.^^J}%
       \usepackage{amssymb}%

      }{}
  \fi
\fi

\ifCUPmtlplainloaded \else
  \IfFileExists{amsbsy.sty}
    {\typeout{^^JFound the 'amsbsy' package on the system, using it.^^J}%
     \usepackage{amsbsy}}
    {\providecommand\boldsymbol[1]{\mbox{\boldmath $##1$}}}
\fi

\usepackage{overpic}
\usepackage{amsmath,mathrsfs}
\usepackage{psfrag}
\usepackage{amsmath,amssymb,amsfonts,bbm}
\usepackage[bbgreekl]{mathbbol}
\usepackage{mathrsfs}
 \usepackage{overpic} 
 \usepackage{hhline}
\usepackage{multirow}
\usepackage{tikz}

\renewcommand{\vec}[1]{\ensuremath{\mbox{\boldmath$#1$}}}
\newcommand{\ma}[1]{\ensuremath{\mathbb{#1}}}

\newcommand{\ve}[1]{\boldsymbol{#1}}

\renewcommand{\epsilon}{\varepsilon}

\begin{document}

\title{Second-order inertial forces and torques on a sphere in a viscous steady  linear flow} 
\shorttitle{Second-order inertial forces and torques on a sphere} 
\author[F. Candelier, R. Mehaddi, B. Mehlig and J. Magnaudet]
{F. Candelier$^1$, R. Mehaddi$^2$, B. Mehlig$^3$ and J. Magnaudet$^4$}
\affiliation{$^1$Aix-Marseille Univ., CNRS, IUSTI, F-13013 Marseille, France\\
$^2$Universit\'e de Lorraine, CNRS, LEMTA, F-54000 Nancy, France\\
$^3$ Department of Physics, Gothenburg University,
 SE-41296 Gothenburg, Sweden\\
$^4$ Institut de M\'ecanique des Fluides de Toulouse (IMFT), Universit\'e de Toulouse, CNRS, INPT, UPS, Toulouse, France}
\date{}
\maketitle
\begin{abstract}
We compute the full set of second-order inertial corrections to the instantaneous force and torque acting on a small spherical rigid particle moving unsteadily in a general steady linear flow. This is achieved by using matched asymptotic expansions and formulating the problem in a coordinate system co-moving with the background flow. Effects of {\color{black}{unsteadiness and}} fluid-velocity gradients are assumed to be small, but to dominate {\color{black}{in the far field}} over those of the velocity difference between the body and fluid, making the results essentially relevant to weakly positively or negatively buoyant particles. The outer solution (which at first order is responsible for the Basset-Boussinesq history force at short time and for shear-induced forces such as the Saffman lift force at long time) is expressed via a flow-dependent tensorial kernel. The second-order inner solution brings a number of different contributions to the force and torque. Some are proportional to the relative translational or angular acceleration between the particle and fluid, while others take the form of products of the rotation/strain rate of the background flow and the relative translational or angular velocity between the particle and fluid. Adding the outer and inner contributions, the known added-mass force or the spin-induced lift force are recovered, and new effects involving the velocity gradients of the background flow are revealed. The resulting force and torque equations provide a rational extension of the classical Basset-Boussinesq-Oseen equation incorporating all first- and second-order fluid inertia effects resulting from both unsteadiness and velocity gradients of the carrying flow. 
\end{abstract}

\section{Introduction}
\label{intro}
Particle-laden flows are ubiquitous in geophysical and engineering contexts \citep{balachandar2010turbulent}. In such flows, the particle dynamics controls how the particles disperse, how they settle, rise or deposit, how they modify the heat/mass transfer efficiency of the carrying flow, at which rate they collide and then possibly fragment or agglomerate, etc.   
Microscopic descriptions of particle suspensions are challenging, especially because \textit{ab initio} direct numerical simulations (DNS) resolving the local flow past many particles are extremely demanding, both in terms of numerical schemes and computer time. \\
\indent An alternative is to use the one-way coupling approximation in which one solves for the fluid motion first, without considering the presence of the particles, and then advances in time an equation of motion for the particles, given a model for the hydrodynamic force and torque acting on them. For very small particles, the force and torque may be obtained by solving the (possibly unsteady) Stokes equation for the disturbance flow caused by the particle in motion, with input data given by the undisturbed fluid velocity field. In this approximation, the advective terms in the Navier-Stokes equation for the disturbance flow are assumed small compared to the viscous term, so that any finite-Reynolds-number effects are neglected. 

This simple approach has two main limitations. First, it considers strictly independent particles, neglecting any interactions within the particle population. This approximation works well for very dilute suspensions, but when two particles come close to each other, their relative motion is affected by hydrodynamic interactions. 
Second, even for a single particle, it is not known in general how corrections to the above zero-Reynolds-number description affect the particle motion. The larger the particles are, the more important these corrections must become. \\ 
\indent The above approximation in which effects of fluid inertia associated with unsteadiness are possibly large while those due to advection are neglected yields a closed-form expression for the total time-dependent force experienced by a small buoyant spherical particle with radius $a$ and mass $m_p$ moving at velocity $\vec{v}_p(t)$ in a uniform flow with velocity $\vec{U}^\infty(t)$. This is the so-called Basset-Boussinesq-Oseen (BBO) equation \citep{boussinesq1885resistance,basset1888treatise,oseen1910uber}. Defining the mass of fluid $m_f$ corresponding to the particle volume, together with the fluid dynamic and kinematic viscosities $\mu$ and $\nu$, this equation predicts that the total force acting on the particle at time $t$ is
\begin{equation}
\vec{f}(t) =\vec{f}_b (t) - 6 \pi \mu a \Bigg\{\:\vec{u}_s(t)  +   \int_0^t \! \frac{1}{\sqrt{\pi\nu (t-\tau)/a^2}} \frac{\mbox{d}\vec{u}_s(\tau)}{\rm{d}\tau} \mbox{d}\tau\Bigg\}-\frac{1}{2} m_f \Big(\frac{\mbox{d} \vec{u}_s}{\mbox{d}t}\Big)(t)  \,,
\label{eq:bbo}
\end{equation}
in which $\vec{u}_s(t)= \vec{v}_p(t) - \vec{U}^{\infty}(t)$ is the slip velocity of the particle with respect to the fluid and
$\vec{f}_b= m_f \frac{\mbox{d} \vec{U}^\infty}{\mbox{d} t}+ (m_p - m_f)\vec{g} $ may be thought of as the generalised buoyancy force acting on the particle, $\vec{g}$ denoting gravity. In the expression for $\vec{f}_b$, the contribution $m_f \frac{\mbox{d} \vec{U}^\infty}{\mbox{d} t}$ results from the nonuniform pressure distribution at the particle surface induced by the acceleration of the undisturbed flow, and is frequently referred to as the `pressure-gradient' force \citep{batchelor1967introduction}. 
The first term within curly braces is the quasi-steady Stokes drag while the second is the well-known Basset-Boussinesq history force resulting from the unsteady viscous diffusion of the disturbance around the particle arising when the slip velocity changes over time. The last term is the so-called added-mass or virtual-mass force. This force results from the no-penetration condition at the particle surface, which constrains the fluid surrounding the particle to react instantaneously if the particle accelerates with respect to the fluid or \textit{vice versa}.
A counterpart of the BBO equation for a spherical bubble at the surface of which the outer fluid obeys a shear-free condition is also available \citep{Gorodtsov1975,Morrison1976,Yang1991}. In this case, the $6\pi$ pre-factor weighting the curly braces becomes $4\pi$ and the $\pi\{\nu(t-\tau)/a^2\}^{-1/2}$ Basset-Boussinesq kernel is changed into $2\exp\{{9\nu(t-\tau)/a^2}\}\text{erfc}\{\sqrt{{9\nu(t-\tau)/a^2}}\}$. This kernel yields a finite contribution to the total force at short time, in contrast with the Basset-Boussinesq kernel which diverges as $t^{-1/2}$. This difference stresses the fact that history effects are much less severe for a shear-free bubble compared to a rigid sphere, thanks to the slip of the outer fluid along the bubble surface.\\
\indent The BBO equation was extended to arbitrary nonuniform flows by \cite{Gatignol1983} and  \cite{Maxey1983}, still neglecting any advective effect in the disturbance equation {\color{black}{(see table \ref{tablelit} for an overview of the different approximations for the force on a small spherical particle that are discussed in this section).}} \cite{Gatignol1983} also obtained the corresponding equation for the hydrodynamic torque, extending the result previously derived in a uniform flow by \cite{Feuillebois1978}. Once advective effects are assumed negligible in the disturbance dynamics, the nonuniformity of the carrying flow manifests itself in two ways. First, since the background velocity varies with the local position, it is necessary to define the slip velocity at the instantaneous position of the particle centre, $\vec{x}_p(t)$, which leads to the definition 
{\color{black}{\begin{equation}
\vec{u}_s(\vec{x}_p(t))=\vec{v}_p(t)-\vec{U}^{\infty}(\vec{x}_p(t))\,. 
\label{usdef}
\end{equation}}}
Computing the disturbance-induced force then reveals that, in addition to $\vec{u}_s$ and $\rm{d}\vec{u}_s/\rm{d}t$, the last three terms on the right-hand side of (\ref{eq:bbo}) involve Fax\'en corrections proportional to $a^2\nabla^2 \ve U^\infty(\ve x_p(t))$. These corrections are significant if the slip velocity is small (nearly neutrally buoyant particles) and the carrying flow varies significantly and nonlinearly at the particle scale. Second, although advective effects are assumed to have no effect on the disturbance flow, they may be significant in the carrying flow. For this reason, they must be taken into account consistently in the `pressure-gradient' contribution involved in the generalized buoyancy force $\vec{f}_b$. A simple reasoning shows that this is achieved by replacing the time variation of the fluid velocity $\frac{\mbox{d} \vec{U}^\infty}{\mbox{d} t}$ with the Lagrangian acceleration $\braket{\frac{\mbox{D} \vec{U}^\infty}{\text{D} t}}$, the $\braket{\cdot}$ symbol denoting the spatial average over the particle volume. If the fluid acceleration may be considered uniform over this volume or if its spatial variations with respect to the particle centre are odd (such as in a linear flow field), $\braket{\frac{\mbox{D} \vec{U}^\infty}{\mbox{D} t}}$ reduces to $\frac{\mbox{D} \vec{U}^\infty}{\mbox{D} t} \Big|_{\vec{x}_p}$, which is the approximation retained by \cite{Maxey1983}. However, to remain consistent with the treatment used for the disturbance, variations of the fluid acceleration within the particle volume must generally be taken into account. The corresponding expansion yields an extra Fax\'en-type correction, as recognized by \cite{Gatignol1983}, but also additional quadratic contributions proportional to $a^2\nabla^2 \ve U^\infty\cdot\nabla \ve U^\infty$ and $a^2\nabla \ve U^\infty\colon\nabla\nabla \ve U^\infty$ \citep{Rallabandi2021}.\\
\indent As already pointed out, (\ref{eq:bbo}) is obtained by assuming that (i) unsteady effects are strong enough for the time-derivative term in the Navier-Stokes equation to be comparable with the viscous term; and (ii) advective effects are negligible throughout the flow. However, it has been proved both theoretically and numerically that the latter assumption breaks down when the disturbance reaches distances of the order of the Oseen length scale $\ell_u=\nu/||\vec{u}_s||$ \citep{sano1981unsteady, Mei1991, lovalenti1993hydrodynamic}. Indeed, whatever the ratio  $a/\ell_u$ (which may be thought of as the particle slip Reynolds number), advective effects cannot be neglected at such a distance, and these effects result in a wake. In this wake, advection being more efficient than viscous diffusion, the disturbance adjusts more quickly to the new slip velocity $\vec{u}_s(t)$, than in the immediate surroundings of the particle, yielding in most cases a $t^{-2}$-long-time decay of the history force instead of the $t^{-1/2}$-prediction resulting from the Basset-Boussinesq kernel. 
Based on these theoretical findings, several authors have proposed approximate extensions of  the BBO equation to particles moving at finite Reynolds number in a uniform but possibly time-dependent flow. 
 \cite{mei1992flow}, \cite{Mei1994} and \cite{Kim1998} designed semi-empirical kernels that recover the correct asymptotic behaviours in the short and long-time limits. Influence of the Reynolds number was incorporated by introducing empirical functions calibrated against DNS results in the kernel, and replacing Stokes' expression for the quasi-steady drag by empirical fits based on the standard drag curve. These attempts have proved successful up to slip Reynolds numbers of several tens, even in configurations far from those in which the empirical functions were calibrated, such as the unsteady rise of CO$_2$ bubbles rapidly dissolving in water \citep{Takemura2004}. \\
\indent Compared to the above picture, extensions of the BBO equation to particles experiencing finite advective effects in nonuniform flows are much less mature, to say the least.  
This does not have severe consequences regarding the prediction of the drag, which is usually only marginally affected by the local fluid-velocity gradients. However, in many configurations, these gradients are known to induce lift components in the hydrodynamic force. For a sphere, this happens every time the slip velocity is not collinear with one of the eigenvectors of the velocity-gradient tensor. Such lift forces strongly affect the particle dynamics in many situations, such as particle deposition and shear-induced migration in wall-bounded shear flows, or the migration of particles toward high-strain regions or vortex cores in turbulent flows, to mention just a few examples. Most studies devoted to these velocity-gradient-induced lift forces considered a steady framework in which the slip velocity is prescribed and does not vary over time. The best known of these studies is that of \cite{saffman1965lift} who examined the case of a spherical particle immersed in a linear shear flow, with a nonzero slip velocity collinear with the background flow. Assuming the shear-based length scale $\ell_s=(\nu/s)^{1/2}$ (with $s$ the shear rate) to be much smaller than the Oseen length $\ell_u$, he employed the technique of matched asymptotic expansions to calculate the dominant contribution to the shear-induced lift force acting on the particle. This seminal work opened the way to a large number of investigations that considered other canonical linear flows or other orientations of the slip velocity with respect to the flow (see \cite{Stone2000} and \cite{candelier2019time} for reviews). The corresponding predictions for the hydrodynamic forces and torques are widely used. However, they are frequently employed well beyond the original framework in which they were derived and are known to be valid. In particular, Saffman's expression for the shear-induced lift force has routinely been added empirically to the right-hand side of (\ref{eq:bbo}) to compute the path of particles immersed in nonuniform time-dependent environments barely resembling a stationary linear shear flow; e.g. \cite{McLaughlin1989} and \cite{li1992dispersion}.  \\
\indent Few studies have attempted to derive a rigorous expression for the total hydrodynamic force acting on particles moving in steady linear flows with a time-varying slip velocity. Their common point is the central assumption that the leading-order solution is governed by the quasi-steady Stokes equation. In other words, effects of unsteadiness are assumed to take place over a characteristic time of the order of the inverse shear rate, $s^{-1}$, allowing the time rate-of-change term to be treated as a perturbation, similar to the shear-induced advective terms. This is in contrast with the assumption on which (\ref{eq:bbo}) is grounded, which considers that velocity variations may take place over a characteristic time possibly as short as the viscous time $a^2/\nu$. With the former assumption, the first-order corrections to the quasi-steady Stokes force arise at order $\epsilon=a/\ell_s$ and include contributions due to the time variations of the slip velocity as well as to the nonlinear advective interaction of the disturbance with the background velocity field. 
\cite{miyazaki1995drag} employed the so-called induced-force method to derive the $\mathcal{O}(\epsilon)$-correction to the force in the case of a linear shear flow, recovering Saffman's prediction in the long-time limit. \cite{asmolov1999inertial} made use of the matched asymptotic expansions technique to obtain the time-dependent lift force in the same configuration, in the specific case of a sphere undergoing periodic oscillations. More recently, \cite{candelier2019time}, {\color{black}{hereinafter referred to as CMM,}} expressed the governing equations for the disturbance in a non-orthogonal coordinate system moving with the undisturbed flow and solved them using matched asymptotic expansions for the three canonical planar linear flows, namely solid-body rotation, uniform straining, and uniform shear. At this point, it is important to stress that the problem is non-linear, owing to the nonlinearity of the advective term. Hence, the solution for an arbitrary linear flow cannot be obtained \textit{via} a linear superposition of the elementary contributions corresponding to solid-body rotation and uniform straining motion \citep{candelier2006analytical}. CMM expressed the $\mathcal{O}(\epsilon)$-correction to the force and torque in the form of a convolution integral involving a tensorial kernel whose components are specific to the linear flow under consideration. This kernel does not depend on the shape of the particle. For this reason, their approach allows the results obtained for a sphere to be straightforwardly extended to arbitrarily shaped particles, simply by performing the dot product of the convolution integral with the appropriate resistance tensor of the particle determined in the creeping-flow limit. {\color{black}{Since effects of unsteadiness are assumed to be small compared to viscous effects, the results obtained through this approach are in general not valid at very short times following the introduction of the particle in the flow, i.e. for times $t\lesssim a^2/\nu$. In contrast, these results are valid in the intermediate range $a^2/\nu\lesssim t\lesssim s^{-1}$ which corresponds to short times with respect to the characteristic time imposed by the velocity gradient.}} At such `short' times, the kernel is diagonal and its nonzero components behave as $t^{-1/2}$, recovering the contribution of the Basset-Boussinesq term in (\ref{eq:bbo}) to the total force and torque. Corrections to this initial behaviour develop gradually over time, both in the diagonal and off-diagonal components. Their evolution depends on the background flow. For instance, the off-diagonal components, responsible for the lift force, grow as $t^{1/2}$ in a uniform shear flow, but only as $t^{5/2}$ in a solid-body rotation flow. Each component eventually converges toward a steady-state value {\color{black}{for $t\gg s^{-1}$}} but the duration of the corresponding transient significantly varies from one component to the other. {\color{black}{At this point, we need to stress that the validity of the BBO equation in the presence of a linear background flow is limited to times usually significantly shorter than $s^{-1}$. For instance, figure 4 in CMM indicates that in a pure shear flow, the lift component which eventually yields the Saffman lift force has already reached $20\%$ (resp. $80\%$) of its final value at $t=\frac{1}{10}s^{-1}$ (resp. $t=s^{-1}$), an effect which is not captured by the BBO approximation. As a consequence, the BBO equation describes for example the dynamics of a spherical particle with radius $a=100\,\mu$m moving in a water flow with $s=1\,$s$^{-1}$ up to $t=a^2/\nu=10^{-2}\,$s, but fails to capture the growth of the shear-induced lift that becomes significant for $t=\frac{1}{10}s^{-1}=10^{-1}\,$s. The situation becomes obviously worse as the shear rate increases, showing that, in this range of particle sizes, the time interval over which the BBO equation is valid does not exceed a few characteristic viscous times.}}

\begin{table}
{\color{black}{ 
  \renewcommand{\arraystretch}{1.5}
 \hspace{-4mm}
  \begin{tabular}{|m{3.2cm}||m{1.9cm}|m{1.9cm}|m{1.9cm}|m{1.8cm}|m{1.8cm}|}
    \hline
    \multirow{8}{2cm}{} & \multicolumn{3}{c|}{\textbf{Unsteady effects in:}} & \multicolumn{2}{c|}{\textbf{Advective effects due to:}}\\
\cline{2-6}
    & \centering\text{uniform} &\centering\text{linear}& \centering\text{arbitrary} & \centering\text{relative} &\centering\text{linear$^{(1)}$}\tabularnewline
      &\centering carrying flow& \centering carrying  flow& \centering carrying  flow &\centering\text{slip}  &\centering\text{carrying flow}\tabularnewline
    \hline\hline
    BBO (eq. (\ref{eq:bbo})) & \centering{\Large$\bullet$}\par\text{$\mathcal{O}(1)$} & & & &\\ \hline
   \cite{Gatignol1983,Maxey1983} &  & & \centering{\Large$\bullet$}\par\text{$\mathcal{O}(1)$}  &  &\\ \hline
    \cite{Kaplun1957,Proudman1957,rubinow1961transverse}$^{(2)}$ & & &  & \centering{\Large$\bullet$} \par\text{$\mathcal{O}(a/\ell_u)$} &\\ \hline
    \cite{sano1981unsteady,Mei1991,lovalenti1993hydrodynamic} &  \centering{\Large$\bullet$} \par\text{$\mathcal{O}(1)$}&  & &\centering{\Large$\bullet$} \par\text{$\mathcal{O}(a/\ell_u)$}  &\\ \hline
    \cite{saffman1965lift, Gotoh90} & &  & &  &\centering{\Large{$\bullet$}} \text{$\mathcal{O}({\epsilon})$}\par \text{$\ell_u\gg\ell_s$}   \tabularnewline  \hline
    \cite{asmolov1990,mclaughlin1991inertial} & &  & &  & \centering{\Large{$\bullet$}} \text{$\mathcal{O}({\epsilon})$}\par \text{$\ell_u/\ell_s=\mathcal{O}(1)$}   \tabularnewline \hline
    \cite{miyazaki1995drag,candelier2019time}& & \centering{\Large{$\bullet$}} \text{$\mathcal{O}({\epsilon})$}\par \text{$\ell_u\gg\ell_s$}  & &  &\centering{\Large{$\bullet$}} \text{$\mathcal{O}({\epsilon})$}\par \text{$\ell_u\gg\ell_s$}   \tabularnewline \hline
    \text{Present work} & & \centering{\Large$\bullet$} \text{$\mathcal{O}({\epsilon,\,\epsilon^2})$} \par \text{$\ell_u\gg\ell_s$}  &  & & \centering{\Large$\bullet$} \text{$\mathcal{O}({\epsilon,\,\epsilon^2})$} \par \text{$\ell_u\gg\ell_s$} \tabularnewline \hline
  \end{tabular}
  \caption{Review of various approximations beyond the quasi-steady Stokes solution for the hydrodynamic force on a small spherical particle. Values in the $\mathcal{O}(\cdot)$ symbols compare the magnitude of unsteady and/or advective effects with that of viscous effects. The ratio $a/\ell_u=a||\vec{u}_s||/\nu$ is the slip Reynolds number based on the relative velocity between the particle and fluid, while $\epsilon=a/\ell_s=(a^2s/\nu)^{1/2}$ is the square root of the shear-based Reynolds number; both Reynolds numbers are assumed small. 
  $^{(1)}$ This includes uniform shear \citep[e.g.][]{saffman1965lift}, solid-body rotation  \citep[e.g.][]{Gotoh90}, and uniform two-dimensional strain; $^{(2)}$ \cite{rubinow1961transverse} considered the combined effect of particle slip and spin.}}}
  \label{tablelit}
\end{table}
\indent The approach outlined above yields a consistent prediction of the hydrodynamic force and torque at $\mathcal{O}(\epsilon)$ in linear flows, provided the slip velocity does not vary `too' rapidly, and advective effects due to the linear background flow dominate over those due to the slip velocity, i.e. $\ell_s\ll\ell_u$. Thus, it may be viewed as the desired rational first-order extension of Stokes' quasi-steady prediction incorporating finite inertial effects, be they due to unsteadiness or to advection by the linear background flow. However, referring to (\ref{eq:bbo}), it is clear that this first-order extension does not include any added mass contribution, owing to the restriction imposed to the time variation of the slip velocity. Such a contribution appears only at next order in the expansion with respect to $\epsilon$.\\
\indent  For reasons of simplicity, many theories for the dynamics of  inertial particles in turbulence just use the quasi-steady Stokes approximation without added-mass or history terms \citep{gustavsson2016statistical}. For similar reasons, and also to limit the computational cost, a number of studies devoted specifically to the dynamics of light particles in turbulence retain the added-mass force but ignore any
$\mathcal{O}(\epsilon)$-contribution, which makes the underlying model formally inconsistent and questions the relevance of its predictions \citep{Bab00,bec2003fractal,Cal08,Cal08b,Cal09,Vol08,vajedi2016inertial}. Added-mass effects are known to be central in the dynamics of light particles, not to mention bubbles. Therefore, determining the $\mathcal{O}(\epsilon^2)$-corrections to the above expansion appears as essential to provide a robust model for studying the dynamics of relatively light particles in which all potentially physically relevant building blocks are included. Similarly, the possible rotation of the particle does not have any influence on the $\mathcal{O}(\epsilon)$-correction to the force and torque. However, it is well known that, as soon as effects of fluid inertia come into play, a spinning particle translating in a fluid at rest experiences a lift force proportional to the cross product of the slip velocity and the spinning rate \citep{rubinow1961transverse}, a clear $\mathcal{O}(\epsilon^2)$-effect. Other cross effects affecting the particle dynamics may be expected at the same order, due to the possible quadratic terms arising from the various combinations of the slip velocity, spinning rate, strain rate and vorticity of the background flow.\\
 \indent When the added-mass force is mentioned, the known expression for the inviscid hydrodynamic force on a sphere in motion in a general linear flow field comes to mind. In present notation, this expression reads \citep{Auton1988}
 \begin{equation}
\vec{f}_{AHP} = \vec{f}_b -  \frac{1}{2} \: m_f \left( \frac{\mbox{d} \vec{v}_p}{\mbox{d}t} - \frac{\mbox{D} \vec{U}^\infty}{\mbox{D} t} \Big|_{\vec{x}_p} \right) - \frac{1}{2} \:m_f\: \vec{u}_s \times \left(\vec{\nabla} \times  \vec{U}^\infty\right)\,,
 \label{auton}
\end{equation}
with $\vec{f}_b=m_f\frac{\mbox{D} \vec{U}^\infty}{\mbox{D} t} \Big|_{\vec{x}_p}+(m_p-m_f)\vec{g}$ in the case of a linear flow. In (\ref{auton}), the disturbance-induced force appears to result from the addition of an added-mass force and a shear-induced lift force. The latter was first computed by \cite{Auton1987} for a sphere held fixed in a stationary uniform shear flow, under the condition that the shear-induced velocity at the body scale is weak compared to the slip velocity, i.e. $\eta=as/||\vec{u}_s||\ll1$. This condition is needed for the velocity correction induced by the distortion of the background vorticity to remain small compared to the slip-induced velocity at the body surface, which in turn allows  the pressure distribution at this surface, hence the force, to be computed at first order with respect to $\eta$. \\
\indent The mathematical form of the added-mass force in (\ref{auton}) involves the Lagrangian acceleration of the background flow, in agreement with the expression established by \cite{taylor1928forces} for pure straining (i.e. irrotational) flows. Noting that  $\frac{\mbox{D} \vec{U}^\infty}{\mbox{D} t} \Big|_{\vec{x}_p}=\frac{\mbox{d} \vec{U}^\infty}{\text{d} t}\Big|_{\vec{x}_p}-\vec{u}_s\cdot\vec{\nabla}\vec{U}^\infty$, {\color{black}{with $\frac{\mbox{d} \vec{U}^\infty}{\text{d} t}\Big|_{\vec{x}_p}$ the time derivative of $\vec{U}_\infty$ following the particle path,}} allows this force, say $\vec{f}_{am}$, to be re-written in the equivalent form 
{\color{black}{\begin{equation}
\vec{f}_{am} = -  \frac{1}{2} \: m_f \left( \frac{\mbox{d} \vec{v}_p}{\mbox{d}t} - \frac{\mbox{D} \vec{U}^\infty}{\mbox{D} t} \Big|_{\vec{x}_p} \right)\equiv -  \frac{1}{2} \: m_f \left( \frac{\mbox{d} \vec{u}_s}{\text{d}t}+\vec{u}_s\cdot\vec{\nabla}\vec{U}^\infty\big|_{\vec{x}_p}\right) \,,
\label{autons}
 \end{equation}}}\noindent
 with the slip velocity evaluated at the particle centre. The latter form of $\vec{f}_{am}$ makes it clear that this force may be nonzero even though the slip velocity does not change over time, provided the carrying flow varies in the direction collinear with the slip. 
According to (\ref{auton}), Taylor's expression applies even if the carrying flow has nonzero vorticity. Moreover, Auton's expression for the shear-induced lift force appears to apply even if the carrying flow is unsteady. However this extension holds only as long as the magnitude of the  slip rate of change, $||\rm{d}\vec{u}_s/\rm{d}t||$, is small compared with $||\vec{u}_s||^2/a$, in which case the stretching/tilting term in the vorticity equation remains primarily balanced by the advective term. The above conditions limit the validity of (\ref{auton}) to weakly unsteady and nonuniform flows, owing to the consequences of the distortion of the upstream vorticity in the three-dimensional flow past a sphere. These limitations do not exist in the two-dimensional case. That is, (\ref{auton}) (with the pre-factors $\frac{1}{2}$ replaced with $1$) is an exact solution for the inviscid force acting on a circular cylinder immersed in a general linear time-dependent flow.\\
 \indent The connection between the inviscid force (\ref{auton}) and the proper generalization of the visco-inertial force (\ref{eq:bbo}) to nonuniform flows has been considered by several authors \citep{Maxey1983,Magnaudet1995}. In particular, \cite{Maxey1983} pointed out that the creeping-flow limit does not allow to decide whether the form of the added-mass term in (\ref{eq:bbo}) remains unchanged when the carrying flow is nonuniform, i.e. whether it still involves the time derivative of the slip velocity following the particle, or whether it rather includes the $-\frac{1}{2}m_f\vec{u}_s\cdot\vec{\nabla}\vec{U}^\infty\big|_{\vec{x}_p}$ contribution resulting from the Lagrangian fluid acceleration in (\ref{autons}). Indeed, this term is a quadratic cross effect in the sense employed above, and consequently appears only at $\mathcal{O}(\epsilon^2)$. Thus, computing the second-order inertial corrections is a necessary step to clarify this issue, as well as to quantify the role of the no-slip condition on the shear-induced lift force by comparing the corresponding prediction with that of (\ref{auton}).
\\
 \indent Given the open questions identified in this introduction, there is a clear need to calculate 
all $\mathcal{O}(\epsilon^2)$-contributions to the force and torque acting on a rigid spherical particle immersed in a stationary linear flow. 
Indeed, as shown above, these contributions contain several physical effects that are not captured at $\mathcal{O}(\epsilon)$, and are thus of fundamental importance to better understand how (possibly coupled) effects of slip, spin, unsteadiness and background velocity gradients modify the simpler picture provided by the $\mathcal{O}(\epsilon^0)$- and $\mathcal{O}(\epsilon)$-models. Obviously, one intuitively expects second-order contributions to only produce small changes in the particle dynamics since $\epsilon$ is assumed small. However, lift effects on a sphere only arise at $\mathcal{O}(\epsilon)$. Therefore, it might be that the sign and magnitude of the $\mathcal{O}(\epsilon)$- and $\mathcal{O}(\epsilon^2)$-contributions to the lift force combine in such a way that the force becomes dominated by second-order effects beyond some critical, but still small, $\epsilon$. Indeed, results discussed later will confirm this possibility in some flows, providing an additional confirmation that these effects are worthy of investigation.\\
\indent 
In the following, we describe how we computed all $O(\epsilon^2)$-corrections to the force and torque on a small rigid sphere moving in a steady linear flow with a time-dependent slip velocity. 
We formulate the problem and the underlying assumptions in \S\,\ref{problem}.
The technique employed to obtain the corresponding contributions makes use of matched asymptotic expansions.  The way the outer problem is solved at the required order is a direct extension of the approach developed by CMM to obtain the $\mathcal{O}(\epsilon)$-corrections. We summarize this approach in \S\,\ref{sec:sol}, show how it extends to order $\mathcal{O}(\epsilon^2)$, and insist on the solution of the inner problem which is required to obtain the complete set of second-order contributions. In  \S\,\ref{disc} we extensively discuss the general results derived in the previous section by considering first the short-term limit {\color{black}{(which in the framework of the present theory corresponds to the intermediate range $a^2/\nu\ll t \ll s^{-1}$)}} for the expressions of the force and torque, then the long-term limit $t\gg s^{-1}$ in four canonical linear flow configurations. We summarize the main outcomes of the study in \S\,\ref{conclu}, providing in particular the complete expressions for the force and torque up to $\mathcal{O}(\epsilon^2)$-terms in both the short- and long-time limits. {\color{black}{Readers mostly interested in applications may directly consider the final results (\ref{statd2})-(\ref{shortd1}), together with the specific expressions of the kernel in the different flow configurations, to obtain a complete view of the effects involved in the force and torque balances at the order of approximation considered here.}}
\begin{figure}
\vspace{-4mm}
\begin{overpic}[scale=0.4,angle=-90]{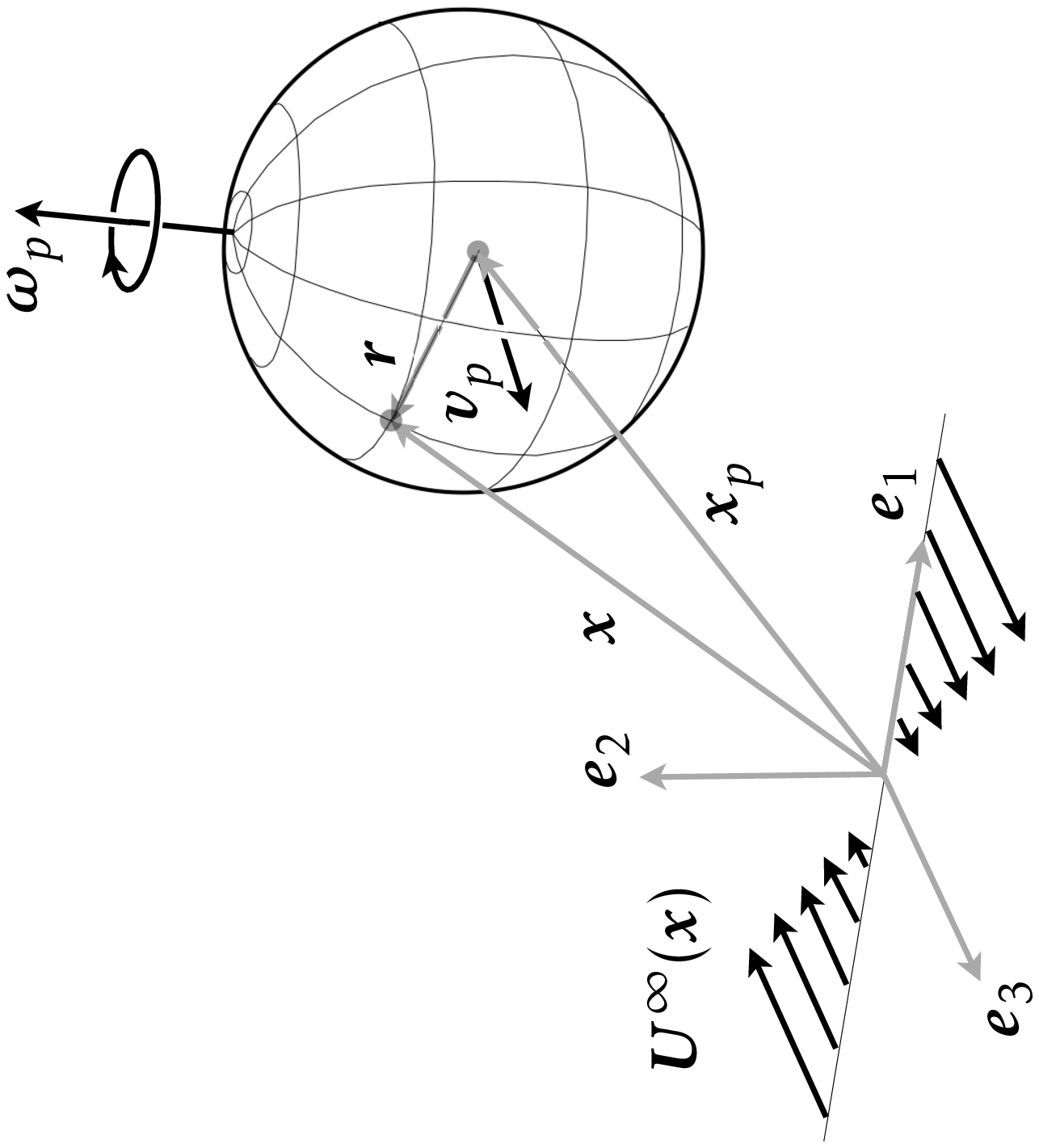}
\end{overpic}
 \label{sketch}
 \vspace{-8mm}
 \caption{
 Sketch of a spherical particle moving in a steady linear flow, with some definitions used throughout the paper.}
 \end{figure}
\section{Basic assumptions and disturbance-flow equations}
\label{problem}
We consider a spherical particle of radius $a$ moving freely in a linear flow of the form
\begin{equation}
\vec{U}^\infty(\vec{x},t) =\vec{U}_0(t)+ \ma{A}\cdot \vec{x} \:,
\label{eq:Uinfty}
\end{equation}
where $\vec{x}$ is the position vector in the laboratory reference frame {\color{black}{whose origin coincides with the position at which the undisturbed flow vanishes}} (see figure \ref{sketch}). {\color{black}{For reasons discussed in CMM and outlined below in \S\,\ref{outerp}, a major simplification is introduced in the calculation of the disturbance induced by the particle in the far field by assuming that $\ma{A}$ does not depend upon time, i.e. the undisturbed flow is stationary. However this puts some restriction on the class of linear flows that can be considered, since $ \ma{A}$ is also uniform. Indeed, combining these two assumptions implies that the balance for the possibly nonzero vorticity $\boldsymbol{\nabla} \times\vec{U}^\infty$ reduces to the stretching term $ \ma{A}\cdot(\boldsymbol{\nabla} \times\vec{U}^\infty)\equiv\ma{S}\cdot(\boldsymbol{\nabla} \times\vec{U}^\infty)$, with $\ma{S} = \tfrac{1}{2}\left(\ma{A} +\ma{A}^{\sf T}\right)$  the strain-rate tensor,  the superscript $^{\sf T}$ denoting the transpose. Consequently, this stretching term must be zero, a constraint satisfied by every planar base flow. In contrast, in three-dimensional configurations, this constraint implies that the general solutions obtained below only hold for purely irrotational base flows.}}\\
\indent Since the undisturbed fluid velocity is linear in $\ve x$, no Fax\' en corrections can be involved in the loads acting on the particle. 
Computing the second-order inertial contributions to the force and torque requires solving
the Navier-Stokes equation for the disturbance flow $\ve w$ caused by the particle motion relative to the undisturbed fluid. To identify the order of magnitude of each term involved in this equation and in the associated boundary conditions, we normalize the fluid and particle translational velocities
with a velocity scale $u_c$, the angular velocities with $\omega_c$, distances with the sphere radius $a$, 
components of $\ma{A}$ with a characteristic strain rate $s$, pressure with $\mu u_c/a$, and time with a characteristic time $\tau_c$ over which the relative translational and angular velocities may vary. 
Using these normalisations, and keeping the previous notations unchanged although all variables are non-dimensional for now on, the equations that govern the disturbance become (CMM)
\begin{subequations}
\label{eq:eom0}
\begin{align}
\boldsymbol{\nabla} \cdot {\vec{w}}& = \vec{0}\:,\\
\label{eq:outer_tp}
\rm{Re}_s \mathrm{Sl} &\frac{ \partial \vec{w}}{\partial t} \Big|_{\vec{r}}
+ \rm{Re}_s \left(  \ma{A} \cdot \vec{w} 
+ (\ma{A} \cdot \vec{r}) \cdot \boldsymbol{\nabla} \vec{w} 
 \right)  +{ \rm Re}_p \left(-\vec{u}_s \cdot\vec{\nabla} \vec{w} + \vec{w}\cdot\vec{\nabla} \vec{w} \right)\nonumber
\\& = - \boldsymbol{\nabla} p + \boldsymbol{\triangle}{\vec{w}} \:,\\
 \quad \vec{w}& \to \vec{0} \quad \mbox{for}\quad r \to \infty\:,
\end{align}
\end{subequations}
with, as shown in figure \ref{sketch}, $\vec{r}=\vec{x}-\vec{x}_p$ the local distance to the particle centre (hence $r=||\vec{r}||=1$ at the particle surface), the time derivative in (\ref{eq:outer_tp}) being evaluated at fixed $\vec{r}$. {\color{black}{According to (\ref{usdef}) and (\ref{eq:Uinfty}), the slip velocity in (\ref{eq:outer_tp}) is $ \vec{u}_s(t)= \vec{v}_p(t)-\ma{A}\cdot \vec{x}_p$. However, the entire problem is left unchanged if, in addition to the linear stationary component $\ma{A}\cdot\vec{x}$, the undisturbed flow is assumed to comprise a uniform time-dependent component, say $\vec{U}_0(t)$, provided the slip velocity is redefined in accordance with  (\ref{usdef}) as $ \vec{u}_s(t)= \vec{v}_p(t)-\vec{U}_0(t)-\ma{A}\cdot \vec{x}_p$.}} For a rigid spherical particle, the no-slip condition at the particle surface implies
\begin{equation}
\label{Eq1_bc}
\vec{w}  = \vec{u}_s + \frac{\mbox{Re}_\omega}{\mbox{Re}_p} \boldsymbol{\omega}_p \times \vec{r}-  \frac{\mbox{Re}_s}{\mbox{Re}_p}\left( \boldsymbol{\Omega} \times \vec{r} + \ma{S} \cdot \vec{r}  \right) \quad \mbox{for}\quad r=1\,,
\end{equation}
with $\boldsymbol{\omega}_p$ the particle angular velocity 
and $ \boldsymbol{\Omega} =\tfrac{1}{2}\vec{\nabla} \times  \vec{U}^\infty$ half the undisturbed flow vorticity, the antisymmetric tensor associated with $\vec{\Omega}$ being the antisymmetric part of $\ma{A}$.  
The non-dimensional parameters in (\ref{eq:eom0}) and (\ref{Eq1_bc}) are the rotation, shear, and slip Reynolds numbers, respectively, plus a Strouhal number characterizing the magnitude of the time-derivative term in (\ref{eq:outer_tp}), namely
\begin{equation}
\mathrm{Re}_\omega = \frac{a^2 \omega_c}{\nu}\:,\quad \mathrm{Re}_s= \frac{a^2 s}{\nu}\:,\quad \mathrm{Re}_p = \frac{a u_c}{\nu} \,,\quad \rm{Sl}=\frac{1}{s \tau_c}\:. 
\end{equation}
To simplify (\ref{eq:eom0})-(\ref{Eq1_bc}), we assume that the particle is small and only weakly positively or negatively buoyant. Therefore, its slip velocity is expected to be small and so is the slip Reynolds number, $\mathrm{Re}_p$.
For the same reason, if the particle is free to rotate, its angular velocity $\boldsymbol{\omega}_p$  is assumed to be close to $\boldsymbol{\Omega}$. We nevertheless keep track of the relative (or slip) angular velocity $\boldsymbol{\omega}_s(t)=\boldsymbol{\omega}_p(t)-\boldsymbol{\Omega}$ in order to allow for a possible transient regime or for a forced particle spin. However, we assume that the magnitude of the particle angular velocity remains of the same order as the shear rate $s$ of the undisturbed flow. {\color{black}{Having defined 
\begin{equation}
\epsilon\equiv\frac{a}{\ell_s}\equiv\mathrm{Re}_s^{1/2}\,,
\end{equation}}}
we therefore set
\begin{subequations}
\label{adim0}
\begin{equation}
u_c\equiv a s\,,\quad\mbox{so that} \quad \mathrm{Re}_p=\epsilon^2 \ll\mathrm{Re}_s^{1/2}\,,
\end{equation}
\begin{equation}
\omega_c \equiv s\quad  \mbox{and} \quad \tau_c \equiv s^{-1}\,,  \quad \mbox{so that} \quad \mathrm{Re}_\omega = \mathrm{Re}_s=\epsilon^2 \quad \mbox{and} \quad \mathrm{Sl}=1\,.
\label{res}
\end{equation}
\\
\end{subequations}
{\color{black}{The assumption $\text{Re}_p\ll\text{Re}_s^{1/2}$ corresponds to the limit considered by Saffman in the stationary $\mathcal{O}(\epsilon)$-problem \citep{saffman1965lift}. It allows effects of slip (but not those due to unsteadiness) to be disregarded in the far field, and we shall show in \S\,\ref{outerp} that this simplification still holds at $\mathcal{O}(\epsilon^2)$. Conditions (\ref{res}) are less critical and simply allow effects of slip, shear, unsteadiness, and spin to all influence the disturbance close to the particle at the retained order of approximation.}} With these assumptions,  the non-dimensional Navier-Stokes equations for the disturbance take the form
\begin{subequations}
\label{eq:eom1}
\begin{equation}
\boldsymbol{\nabla} \cdot {\vec{w}} = \vec{0}\:,
\label{eqs_cont}
\end{equation}
\begin{equation}
\epsilon^2 \frac{ \partial \vec{w}}{\partial t} 
+ \epsilon^2 \left[  \ma{A}\cdot \vec{w} 
+ (\ma{A}\cdot \vec{r}) \cdot \boldsymbol{\nabla} \vec{w} 
 \right]  + \epsilon^2 \left(-\vec{u}_s \cdot\vec{\nabla} \vec{w} + \vec{w}\cdot\vec{\nabla} \vec{w} \right)
 = - \boldsymbol{\nabla} {p}  + \boldsymbol{\triangle}{\vec{w}} \:,
\label{eqs_final}
\end{equation}
\begin{align}
\vec{w}  &= \vec{u}_s(t) + \left[\boldsymbol{\omega}_s(t) \times \vec{r} - \ma{S}\cdot \vec{r}  \right]\quad\mbox{for} \quad r=1\:, \label{eqs_bca}\\
 \vec{w} &\to \vec{0} \quad\mbox{for}\quad r \to \infty\:.
\label{eqs_bc}
\end{align}
\end{subequations}
{\color{black}{The disturbance is assumed to be zero for $t<0$. To avoid singularities in the force and torque at $t=0$, we assume that the particle is introduced in the flow with zero slip velocity and relative rotation rate ($\vec{u}_s(0)=\vec{0},\,\boldsymbol{\omega}_s(0)=\vec{0}$), but let both quantities have arbitrary nonzero initial time derivatives ($\frac{\mbox{\small{d}}\vec{u}_s}{\mbox{\small{d}}t}(0)\neq\vec{0},\,\frac{\mbox{\small{d}}\boldsymbol{\omega}_s}{\mbox{\small{d}}t}(0)\neq\vec{0}$). In the usual case of freely-moving particles, the subsequent evolution of $\vec{u}_s$ and $\boldsymbol{\omega}_s$ is dictated by the overall force and torque balances on the particles. Here in contrast, we are interested in predicting the hydrodynamic loads resulting from arbitrary evolutions of both quantities. The expressions for the force and torque obtained in this way may then be inserted in the overall force and torque balances, which usually involve nonzero external forces (such as the generalised buoyancy force $\vec{f}_b$ introduced in \S\,\ref{intro}) and possibly external torques, and the actual evolution of $\vec{u}_s$ and $\boldsymbol\omega_s$ ensues.}} Starting from (\ref{eq:eom1}), CMM computed the force and torque to $\mathcal{O}(\epsilon)$. They disregarded terms that
do not contribute to the loads at this order. This includes the {\color{black}{term within square brackets in the right-hand side of}} (\ref{eqs_bca}) because its contribution to the disturbance flow (through a rotlet and a stresslet) vanishes at this order for symmetry reasons, and the advective term due to the slip velocity in (\ref{eqs_final}) which is negligible compared to that due to shear in the far field in the Saffman limit.
However, this term contributes to the inner solution at $\mathcal{O}(\epsilon^2)$.

Here our goal is to determine all $\mathcal{O}(\epsilon^2)$-corrections to the force and torque. When commenting on the corresponding results, we shall frequently refer to the original BBO equation (\ref{eq:bbo}). 
In the non-dimensional variables defined above, this equation reads
\begin{equation}
\vec{f}_{BBO}= \vec{f}_b- 6 \pi \vec{u}_s 
  -   6 \pi \epsilon\int_0^t \frac{1}{\sqrt{\pi (t-\tau)}} \frac{\mbox{d}\vec{u}_s}{\rm{d}\tau}  \mbox{d}\tau -  \frac{2\pi}{3} \epsilon^2 \frac{\mbox{d} \vec{u}_s}{\mbox{d}t}  \:.
  \label{BBO}
\end{equation}
The generalised buoyancy force $\vec{f}_b$ usually comprises an $\mathcal{O}(1)$-contribution due to gravity/buoyancy, complemented with contributions due to the Lagrangian acceleration of the undisturbed flow. 
{\color{black}{Noting that $\frac{\mbox{d} \vec{U}^\infty}{\text{d} t}\Big|_{\vec{x}_p}\equiv\vec{v}_p\cdot\vec{\nabla}\vec{U}^\infty+\dot{\vec{U}}_0$, with $\dot{\vec{U}}_0$ the time derivative of the possible uniform component $\vec{U}_0(t)$ of $\vec{U}_\infty$ evaluated in the laboratory frame, one has $\frac{\mbox{D} \vec{U}^\infty}{\mbox{D} t} \Big|_{\vec{x}_p}=\dot{\vec{U}}_0+(\vec{v}_p-\vec{u}_s)\cdot\vec{\nabla}\vec{U}^\infty$.}} Since the particle velocity may be much larger than the slip velocity, one concludes that the $\frac{\mbox{d} \vec{U}^\infty}{\text{d} t}\Big|_{\vec{x}_p}$ term in the Lagrangian acceleration may contribute up to $\mathcal{O}(1)$ to the total force, whereas the contribution proportional to $-\vec{u}_s\cdot\nabla\vec{U}^\infty$ is of $\mathcal{O}(\epsilon^2)$. Later, we also compare the predictions for the second-order force with their counterpart in the inviscid limit. 
In non-dimensional form, (\ref{auton}) becomes $\vec{f}_{AHP} = \vec{f}_b+\epsilon^2\vec{f}'_{inv}$, where the disturbance-induced force $\vec{f}'_{inv}$ reads
\begin{equation}
\vec{f}'_{inv} = - \frac{2\pi}{3} \Bigg\{ \left( \frac{\mbox{d} \vec{u}_s}{\mbox{d}t} + \vec{u}_s\cdot\nabla\vec{U}^\infty\big|_{\vec{x}_p} \right)+\: \vec{u}_s \times \left(\vec{\nabla} \times  \vec{U}^\infty\right)\Bigg\}\:.
 \label{auton2}
\end{equation}
In contrast to (\ref{BBO}), no contribution arises in the disturbance-induced force at $\mathcal{O}(\epsilon^0)$ and $\mathcal{O}(\epsilon)$ in the inviscid limit. This is because no vorticity is generated at the surface of the sphere, leading in particular to zero viscous drag (D'Alembert paradox). Keeping in mind that the Saffman lift force is an $\mathcal{O}(\epsilon)$-effect in the regime of interest here, the fact that no term exists at this order in (\ref{auton2})  implies that the inviscid shear lift force (last term in the right-hand side) is one order of magnitude smaller than the dominant lift force present in the low-but-finite Reynolds number regime.

\section{Method and solution}
\label{sec:sol}
We use matched asymptotic expansions in the spirit of \cite{childress1964slow} and \cite{saffman1965lift} to approximate the solution of (\ref{eq:eom1}) up to $\mathcal{O}(\epsilon^2)$.
The $\mathcal{O}(\epsilon)$-terms listed below were computed earlier by CMM. Most of the technical steps described in \S\,\ref{outerp} also follow closely the approach developed in that reference.

\subsection{Outer problem}
\label{outerp}
Far from the particle, the disturbance-flow equations (\ref{eq:eom1}) in the outer region simplify thanks to Saffman's assumption, $\rm{Re}_p \ll \epsilon\ll 1$.
Indeed,  the magnitude of the disturbance velocity $\ve w$ scales as $1/r$ for large $r$. In the matching region $r\sim \epsilon^{-1}$, one then has the following estimates
\begin{equation}
-\vec{u}_s \cdot\vec{\nabla} \vec{w} \sim \mathcal{O}(\epsilon^2), \quad\quad 
 \vec{w}\cdot\vec{\nabla} \vec{w}  \sim \mathcal{O}(\epsilon^3) \:,\quad  \quad 
\frac{ \partial \vec{w}}{\partial t} \sim 
\ma{A}\cdot \vec{w} 
+ (\ma{A}\cdot \vec{r}) \cdot \boldsymbol{\nabla} \vec{w}  \sim 
\mathcal{O}(\epsilon)\:.
\label{estim}
\end{equation}
Therefore the first two terms  can be dropped in (\ref{eqs_final}) for $r \sim \epsilon^{-1}$ and beyond, 
yielding the leading-order momentum equation in the outer region
\begin{equation}
\begin{split}
\epsilon^2 
\left( 
\frac{ \partial \vec{w}_{\mbox{\scriptsize out}}}{\partial t} 
+ \ma{A}\cdot \vec{w}_{\mbox{\scriptsize out}} 
+ (\ma{A}\cdot \vec{r}) \cdot \boldsymbol{\nabla} \vec{w}_{\mbox{\scriptsize out}}
 \right) = - \boldsymbol{\nabla} {p}_{\mbox{\scriptsize out}}  
 + \boldsymbol{\triangle}{\vec{w}_{\mbox{\scriptsize out}}}
 +  \left(\vec{f}^{(0)} + \epsilon \vec{f}^{(1)}\right) \delta(\vec{r}) \:.
 \end{split}
\label{eq:outer_Linear_flow}
\end{equation}
In the standard fashion pioneered by \cite{childress1964slow}, we account for the particle  in the outer region through a delta function, $\delta(\ve r)$. Usually, the strength of the corresponding force, $\vec{f}^{(0)}$, is taken to be the opposite of the disturbance-induced force experienced by the particle in the creeping-flow limit, $\vec{f}'^{(0)}$. Here, however, we must allow for corrections of $\mathcal{O}(\epsilon)$ since we wish to compute 
the outer contribution to the force at $\mathcal{O}(\epsilon^2)$. This is why there is a contribution $\epsilon \vec{f}^{(1)}$ on the right hand-side of (\ref{eq:outer_Linear_flow}). Strictly speaking, the right-hand side of (\ref{eq:outer_Linear_flow}) should also comprise an additional source term in the form of a dipolar contribution, $\ma{D}^{(0)}\cdot\ \boldsymbol\nabla\delta(\vec{r})$, resulting from the strain- and rotation-induced terms in the boundary condition (\ref{eqs_bca}). This dipolar term adds a linear correction in the far-field flow, which may be computed after the components of the second-rank tensor $\ma{D}^{(0)}$ have been evaluated by matching the leading-order inner and outer flows in the intermediate region $r\sim\epsilon^{-1}$. This matching procedure yields $\ma{D}^{(0)}\cdot\ \boldsymbol\nabla\delta = \frac{20 \pi}{3} \ma{S}\cdot\ \boldsymbol\nabla\delta - 4\pi \boldsymbol{\omega}_s\times \boldsymbol\nabla\delta$, which corresponds to the stresslet and torque exerted by the particle on the fluid \citep{Batchelor1970}. Nevertheless, the corresponding far-field correction does not change the force on the particle at any order. Moreover, it may be shown that it only modifies the torque at $\mathcal{O}(\epsilon^3)$ \citep{meibohm2016angular}. For these reasons, we ignore this dipolar term in what follows. {\color{black}{We also need to examine whether or not the Oseen-like contribution resulting from the first-order disturbance in the far field, $-\vec{u}_s\cdot\boldsymbol{\nabla}\vec{w}_{\mbox{\scriptsize out}}^{(1)}$, has to be included in (\ref{eq:outer_Linear_flow}) to evaluate the second-order correction, $\vec{w}_{\mbox{\scriptsize out}}^{(2)}$. For reasons detailed below, it turns out that this additional forcing term does not add any singular correction to the far-field flow. Therefore, it is sufficient to consider the effect of Oseen-like contributions upon the inner solution to evaluate the modification they introduce on the loads experienced by the particle. This is why no Oseen-like term is included in (\ref{eq:outer_Linear_flow}). We shall return to this point later in this section.}}\\
\indent Equation (\ref{eq:outer_Linear_flow}) can be solved via Fourier transform (see CMM). In Fourier space, {\color{black}{once the pressure has been eliminated with the aid of the continuity equation,}}
the transformed outer momentum equation reads
\begin{equation}
\begin{split}
\epsilon^2 
\left( 
\frac{ \partial \hat{\vec{w}}_{\mbox{\scriptsize out}}}{\partial t}\Big|_{\vec{k}} 
+ \ma{A}\cdot \hat{\vec{w}}_{\mbox{\scriptsize out}} 
- \vec{k} \cdot \ma{A}\cdot \hat{\vec{\nabla}} \hat{\vec{w}}_{\mbox{\scriptsize out}} -2  \left[\frac{ (\ma{A}\cdot \hat{\vec{w}}_{\mbox{\scriptsize out}}) \cdot \vec{k} }{k^2}\right] \vec{k}  \right) =  \\  
-  k^2 \hat{\vec{w}}_{\mbox{\scriptsize out}}  +  k^2 \:\hat{\ma{G}} \cdot  \left(\vec{f}^{(0)} + \epsilon \vec{f}^{(1)}\right) \:,
\end{split}
\label{eq:outer_Linear_flow_Fourier2}
\end{equation}
where 
\begin{equation}
\hat{\ma{G}} = \frac{1}{k^2}\left(\ma{I} - \frac{\vec{k} \otimes \vec{k}}{k^2}\right)
\end{equation}
is the Fourier transform of the Green function $\ma{G}(\vec{r})$ of the Stokes equation, with $\ma{I}$ the unit matrix. {\color{black}{Following CMM, (\ref{eq:outer_Linear_flow_Fourier2}) is expressed in a non-orthogonal coordinate system that moves and deforms with the background flow. This transformation allows the problem to be reduced to a set of ordinary differential equations that are much easier to solve. However this simplification only holds as long as $\ma{A}$ does not depend upon time. Indeed, if $\ma{A}$ is time-dependent, an extra term arises in the transformed momentum equation, complicating the  structure of the general solution and making it much more difficult to obtain. This is why the outer solution described below is only valid when the linear flow is stationary.\\}}
\indent Once $\hat{{\vec{w}}}_{\mbox{\scriptsize out}}$ is obtained, it may be expanded {\color{black}{in terms of generalized functions of $\vec{k}$}} in the form \citep{meibohm2016angular}
\begin{equation}
\hat{{\vec{w}}}_{\mbox{\scriptsize out}}  = \widehat{\vec{\mathcal{T}}}_{\mbox{\scriptsize reg}}^{(0)}  + \epsilon\big( \widehat{\vec{\mathcal{T}}}_{\mbox{\scriptsize reg}}^{(1)} + \widehat{\vec{\mathcal{T}}}_{\mbox{\scriptsize sing}}^{(1)}\big)  + \epsilon^2 \big(\widehat{\vec{\mathcal{T}}}_{\mbox{\scriptsize reg}}^{(2)} + \widehat{\vec{\mathcal{T}}}_{\mbox{\scriptsize sing}}^{(2)}\big)+\ldots \:
\label{eq_expansion_lin}
\end{equation}
Inserting this ansatz into (\ref{eq:outer_Linear_flow_Fourier2})  and expanding in powers of $\epsilon$ provides
the regular contributions $\widehat{\vec{\mathcal{T}}}_{\mbox{\scriptsize reg}}^{(n)}$, namely
\begin{subequations}
\label{T0_T1reglin}
\begin{align}
\widehat{\vec{\mathcal{T}}}_{\mbox{\scriptsize reg}}^{(0)}  =& \hat{\ma{G}} \cdot  \vec{f}^{(0)}  \:,\\
\widehat{\vec{\mathcal{T}}}_{\mbox{\scriptsize reg}}^{(1)}  =& \hat{\ma{G}} \cdot  \vec{f}^{(1)}  
\:,\\
\widehat{\vec{\mathcal{T}}}_{\mbox{\scriptsize reg}}^{(2)}   = &-\frac{1}{k^2}\hat{\ma{G}}\cdot 
\frac{\mbox{d}\vec{f}^{(0)}}{\mbox{d}t} - \frac{1}{k^4}  \ma{A}\cdot 
\vec{f}^{(0)} \\ \nonumber
&
- \frac{1}{k^6}\Big( 2 [ (\ma{A}\cdot \vec{k}) \cdot \vec{k}] \: \ma{I} - (\ma{A}-\ma{A}^{\sf T}) \cdot \vec{k}\otimes\vec{k} - \vec{k}\otimes\vec{k} \cdot \ma{A}\Big) \cdot \vec{f}^{(0)} \\
&+ \frac{2}{k^8} [(\ma{A} \cdot \vec{k}) \cdot \vec{k} ] \:\vec{k}\otimes\vec{k}  \cdot \vec{f}^{(0)}  \:.\nonumber
\end{align} 
\end{subequations}
Transforming these contributions from Fourier space back to the configuration space
yields
\begin{subequations}
\label{T2regphy}
\begin{align}
&{\vec{\mathcal{T}}}_{\mbox{\scriptsize reg}}^{(0)}  = {\ma{G}} \cdot  \vec{f}^{(0)} \:,\\
& {\vec{\mathcal{T}}}_{\mbox{\scriptsize reg}}^{(1)}  = {\ma{G}} \cdot  \vec{f}^{(1)} 
\:,\\
\nonumber
&{\vec{\mathcal{T}}}_{\mbox{\scriptsize reg}}^{(2)}   =  \frac{3 r}{32\pi} \left( \ma{I} - \frac{1}{3} \frac{\vec{r} \otimes \vec{r}}{r^2}\right) \cdot \frac{\mbox{d} \vec{f}^{(0)}}{\mbox{d}t} \\
& \nonumber  + \frac{r }{32\pi} \left( 3 \left( \ma{I} - \frac{1}{3} \frac{\vec{r} \otimes \vec{r}}{r^2}\right) \cdot \ma{A}    -  (\ma{A}-\ma{A}^{\sf T}) \cdot\left( \ma{I} + \frac{\vec{r} \otimes \vec{r}}{r^2} \right)  + 2\left[\frac{\ma{A}: \vec{r} \otimes \vec{r}}{r^2}\right] \ma{I}\right) \cdot \vec{f}^{(0)} \\
& - \frac{r}{96\pi } \left( 2\ma{S}
+ 2\: \frac{\vec{r} \otimes \vec{r}}{r^2} \cdot \ma{S}
  + 2\:\ma{S} \cdot  \frac{\vec{r} \otimes \vec{r}}{r^2}
   + \left[\frac{\ma{A} : \vec{r} \otimes \vec{r}}{r^2} \right]   \left(\ma{I} - \frac{\vec{r} \otimes \vec{r}}{r^2} \right) \right) \cdot\vec{f}^{(0)} \:,
   \label{inner2}
\end{align}
\end{subequations}
with
\begin{equation}
\ma{G}(\vec{r})= \frac{1}{8\pi} \left( \frac{\ma{I}}{r} + \frac{\vec{r}\otimes\vec{r}}{r^3}\right)\,.
\end{equation}
The expansion (\ref{eq_expansion_lin}) suggests that there are singular terms in $\ve{k}$-space, $\widehat{\vec{\mathcal{T}}}_{\mbox{\scriptsize sing}}^{(n)}$,  that cannot be obtained in this way. {\color{black}{These terms, which are concentrated at $\vec{k}=\vec{0}$, correspond to uniform flow corrections in the far field, resulting from the presence of the particle.}} As is well known, these corrections are directly related to the $1/r$-divergence of the advective contribution $\ma{A}\cdot \vec{w} + (\ma{A}\cdot \vec{r}) \cdot \boldsymbol{\nabla} \vec{w}$ based on the leading-order solution. These singular terms may be computed by evaluating {\color{black}{in the sense of generalized functions}} the successive limits
\citep{meibohm2016angular}
\begin{equation}
 \widehat{\mathcal{T}}_{\mbox{\scriptsize sing}}^{(1)}   =  \lim_{\epsilon \to 0} \frac{\widehat{{\vec{w}}}_{\mbox{\scriptsize out}}  - \widehat{\mathcal{T}}_{\mbox{\scriptsize reg}}^{(0)}}{\epsilon} - \widehat{\mathcal{T}}_{\mbox{\scriptsize reg}}^{(1)}\,,
 \label{eqT1sing}
\end{equation}
and 
\begin{equation}
 \widehat{\mathcal{T}}_{\mbox{\scriptsize sing}}^{(2)}   =  \lim_{\epsilon \to 0} \frac{\widehat{{\vec{w}}}_{\mbox{\scriptsize out}}  - \widehat{\mathcal{T}}_{\mbox{\scriptsize reg}}^{(0)} - \epsilon \widehat{\mathcal{T}}_{\mbox{\scriptsize reg}}^{(1)} -\epsilon \widehat{\mathcal{T}}_{\mbox{\scriptsize sing}}^{(1)}}{\epsilon^2} - \widehat{\mathcal{T}}_{\mbox{\scriptsize reg}}^{(2)} \:.
 \label{eqT2sing}
\end{equation}
The outer solution is obtained from (\ref{eq:outer_Linear_flow_Fourier2}), allowing the successive singular contributions to be evaluated. $\widehat{\vec{\mathcal{T}}}_{\mbox{\scriptsize sing}}^{(1)}$ was calculated in this way by CMM who showed that {\color{black}{it takes the form of a convolution product between a tensorial kernel, $\ma{K}(t)$, and the time derivative of the leading-order forcing term  in (\ref{eq:outer_Linear_flow_Fourier2}), $\vec{f}^{(0)}$. That is,}}
\begin{equation}
\widehat{\vec{\mathcal{T}}}_{\mbox{\scriptsize sing}}^{(1)}   = - 8 \pi^3\left[ \int_0^{t} \ma{K}(t-\tau) \cdot \frac{\mbox{d} \vec{f}^{(0)}}{\mbox{d}\tau} \mbox{d} \tau \right] \delta(\vec{k}) 
\equiv 8 \pi^3  \:\vec{\mathcal{U}}_1(t) \delta(\vec{k}) \:.
\label{T1hat}
\end{equation}
{\color{black}{The $[\ma K]_i^j(t)$ component of the kernel expresses how  an instantaneous change in the $i^{th}$-component of the slip velocity (hence in that of the force exerted by the particle on the fluid) influences the $j^{th}$-component of the uniform velocity correction in the far field at later time. The characteristic time scale over which this influence manifests itself is that imposed by the shear, which is larger than the viscous time scale by a factor of $\mathcal{O}(\epsilon^{-2})$. Since $\widehat{\vec{\mathcal{T}}}_{\mbox{\scriptsize sing}}^{(1)}$ contributes to $\hat{\vec{w}}_{\mbox{\scriptsize out}} $ at $\mathcal{O}(\epsilon)$ according to (\ref{eq_expansion_lin}), it may be concluded that time variations in the slip velocity (hence in $\vec{f}^{(0)}$) already affect the force on the particle at $\mathcal{O}(\epsilon)$ through the far-field velocity correction $\vec{\mathcal{U}}_1(t) $. Conversely, as (\ref{inner2}) shows, these time variations only affect the inner solution at $\mathcal{O}(\epsilon^2)$.}} CMM computed the kernel $\ma K(t)$ and discussed its properties for the three canonical planar linear flows. At very short time, {\color{black}{comparable with the viscous time scale ($t\sim\epsilon^2$),}} or in the absence of fluid-velocity gradients, $\ma K$ reduces to  the Basset-Boussinesq kernel, i.e. $\ma K(t)\rightarrow(\pi t)^{-1/2}\ma{I}$.
The second-order singular term takes the form
\begin{equation}
\widehat{\vec{\mathcal{T}}}_{\mbox{\scriptsize sing}}^{(2)} = -8 \pi^3  \left[ \int_0^{t} \ma{K}(t-\tau) \cdot \frac{\mbox{d} \vec{f}^{(1)}}{\mbox{d}\tau} \mbox{d} \tau \right] \delta(\vec{k})
\equiv 8 \pi^3 \: \vec{\mathcal{U}}_2(t) \delta(\vec{k}) \:.
\label{T2hat}
\end{equation}
Remarkably, this term involves the same kernel, $\ma K(t)$, as $\widehat{\vec{\mathcal{T}}}_{\mbox{\scriptsize sing}}^{(1)}$. Since $\vec{f}^{(1)}(t)=-6\pi \vec{\mathcal{U}}_1(t)$ results from the convolution product between the kernel $\ma{K}(t)$ and the time derivative of the force $\vec{f}^{(0)}$, $\widehat{\vec{\mathcal{T}}}_{\mbox{\scriptsize sing}}^{(2)} $ has the form of a double
convolution product.  
In the configuration space, the above two singular terms give rise to two spatially uniform but time-dependent velocity corrections, namely
\begin{equation}
{\vec{\mathcal{T}}}_{\mbox{\scriptsize sing}}^{(1)} = \vec{\mathcal{U}}_1(t) 
\quad \mbox{and} \quad {\vec{\mathcal{T}}}_{\mbox{\scriptsize sing}}^{(2)} = \vec{\mathcal{U}}_2(t)\:.
\label{T0T1singphy}
\end{equation}
Equations  (\ref{T2regphy}) and (\ref{T0T1singphy}) are the desired solutions of the outer problem. The velocity field resulting from the superposition of these solutions serves as the outer boundary condition for the inner problem, since the inner and outer solutions must match at $r \sim \epsilon^{-1}$. {\color{black}{Note that we also evaluated ${\vec{\mathcal{T}}}_{\mbox{\scriptsize sing}}^{(2)}$ with the Oseen-like term included in (\ref{eq:outer_Linear_flow}). The far-field disturbance resulting from this term comprises odd and even functions of $\vec{k}$, but the former is one order lower with respect to $\epsilon$ than the latter, owing to the assumption that the Saffman length is shorter than the Oseen length by a factor of $\mathcal{O}(\epsilon)$ (see appendix A in CMM for details on the evaluation of the contributions to $\vec{w}_{\mbox{\scriptsize out}}$ in $\vec{k}$-space). Conversely, and for the same reason, the even part is one order lower with respect to $\epsilon$ than the odd one in the case of the disturbance associated with the linear flow. Since odd terms eventually integrate to zero, it turns out that, at the order of approximation we need to consider, the Oseen-like term does not contribute to ${\vec{\mathcal{T}}}_{\mbox{\scriptsize sing}}^{(2)}$ whereas terms associated with the linear flow do. In contrast, the Oseen-like term produces contributions in the regular term ${\vec{\mathcal{T}}}_{\mbox{\scriptsize reg}}^{(2)}$. However, the evaluation of ${\vec{\mathcal{T}}}_{\mbox{\scriptsize reg}}^{(2)}$ is not needed, since this term merely matches the second-order inner solution for $r\sim\epsilon^{-1}$. Therefore, only the contributions induced in this inner solution by the Oseen-like term $-\vec{u}_s\cdot\boldsymbol{\nabla}\vec{w}_{\mbox{\scriptsize in}}^{(1)}$ and the companion term $\vec{w}_{\mbox{\scriptsize in}}^{(1)}\cdot\boldsymbol{\nabla}\vec{w}_{\mbox{\scriptsize in}}^{(1)}$ need to be considered (see below) to evaluate the loads on the body at $\mathcal{O}(\epsilon^2)$.}}

\subsection{Inner problem}
\label{innerp}
At order $\epsilon^0$, the problem to solve  is
\begin{equation}
\boldsymbol{\nabla} \cdot  {\vec{w}}_{\mbox{\scriptsize in}}^{(0)}= \vec{0}\,,
\quad 
  - \boldsymbol{\nabla} {p}_{\mbox{\scriptsize in}}^{(0)} + \boldsymbol{\triangle}  
  {\vec{w}}_{\mbox{\scriptsize in}}^{(0)}  =  \vec{0}\,,
\end{equation}
\begin{equation}
{\vec{w}}_{\mbox{\scriptsize in}}^{(0)} =   \vec{u}_s +\boldsymbol{\omega}_s\times \vec{r} - \ma{S}\cdot \vec{r}  \:, \quad \mbox{at}\quad r=1\,,
\quad 
\quad {\vec{w}}_{\mbox{\scriptsize in}}^{(0)} \sim  \vec{\mathcal{T}}^{(0)}  = \ma{G} \cdot \vec{f}^{(0)}\,, \quad\mbox{for}\quad r \to \infty\,.
\vspace{2mm}
\end{equation}
 In a linear flow, the solution of this standard Stokes problem is known to be \citep{Kim1991}
\begin{equation}
{\vec{w}}_{\mbox{\scriptsize in}}^{(0)} =  \ma{G} \cdot \vec{f}^{(0)} + \frac{1}{6}\vec{\triangle} \Big( \ma{G} \cdot \vec{f}^{(0)} \Big) + \frac{\boldsymbol{\omega}_s\times \vec{r}}{r^3} -\frac{\ma{S}\cdot \vec{r}}{r^5}-\frac{5}{2\:r^5}\left(1-\frac{1}{r^2}\right) \vec{r}\otimes\vec{r}\cdot \ma{S} \cdot \vec{r} \:.
\label{O0}
\end{equation}
Only the first term on the right-hand side contributes to the disturbance-induced force $\vec{f}'^{(0)}=-\vec{f}^{(0)}$ on the particle, which is of course nothing but the Stokes drag corresponding to the slip velocity $\vec{u}_s$, namely
\begin{equation}
\vec{f}'^{(0)}(t) = -6\pi \: \vec{u}_s(t)\:. 
\label{eq:f0}
\end{equation}
Similarly, only the rotlet (third term on the right-hand side of (\ref{O0})) contributes to the torque for a sphere, leading to
\begin{equation}
\vec{\tau}'^{(0)}(t) = -8\pi \boldsymbol{\omega}_s(t)\,. 
\label{eq:t0}
\end{equation}
At order $\epsilon$, the problem to solve is  
\begin{equation}
\boldsymbol{\nabla} \cdot  {\vec{w}}_{\mbox{\scriptsize in}}^{(1)}= \vec{0}\:,
\quad 
  - \boldsymbol{\nabla} {p}_{\mbox{\scriptsize in}}^{(1)} + \boldsymbol{\triangle}  
  {\vec{w}}_{\mbox{\scriptsize in}}^{(1)}  =  \vec{0}\,,
\end{equation}
subject to the boundary conditions
\begin{equation}
{\vec{w}}_{\mbox{\scriptsize in}}^{(1)} =   \vec{0}  \quad \mbox{at}\quad r=1\,, 
\quad 
\quad {\vec{w}}_{\mbox{\scriptsize in}}^{(1)} \sim  \vec{\mathcal{U}}^{(1)}(t)+\vec{\mathcal{T}}^{(1)}_{\rm{reg}} \quad \mbox{for}\quad r \to \infty\:. 
\end{equation}
This is a homogeneous Stokes problem. Its solution may be sought in the form 
\begin{equation}
{\vec{w}}_{\mbox{\scriptsize in}}^{(1)}  = \vec{\mathcal{U}}^{(1)}(t) + {\vec{v}}_{\mbox{\scriptsize in}}^{(1)} \:.
\end{equation}
Since $\vec{\mathcal{U}}^{(1)}(t)$ is uniform (see (\ref{T0T1singphy})), the  problem for ${\vec{v}}_{\mbox{\scriptsize in}}^{(1)}$ becomes
\begin{equation}
\boldsymbol{\nabla} \cdot  {\vec{v}}_{\mbox{\scriptsize in}}^{(1)}= \vec{0}\:,
\quad 
  - \boldsymbol{\nabla} {p}_{\mbox{\scriptsize in}}^{(1)} + \boldsymbol{\triangle}  
  {\vec{v}}_{\mbox{\scriptsize in}}^{(1)}  =  \vec{0}\,,
\end{equation}
\begin{equation}
{\vec{v}}_{\mbox{\scriptsize in}}^{(1)} =    -\vec{\mathcal{U}}^{(1)}(t) \quad \mbox{at}\quad r=1\,, 
\quad{\vec{v}}_{\mbox{\scriptsize in}}^{(1)}  \sim \ma{G} \cdot \vec{f}^{(1)} \quad \mbox{for}\quad r\to \infty\:.  
\label{BC:v_in}
\vspace{2mm}
\end{equation}
This problem is formally identical to the Stokes problem for a sphere moving in a fluid at rest. Therefore the solution for ${\vec{w}}_{\mbox{\scriptsize in}}^{(1)}$ is readily obtained as
\begin{equation}
{\vec{w}}_{\mbox{\scriptsize in}}^{(1)} =\vec{\mathcal{U}}^{(1)}(t) + \ma{G} \cdot \vec{f}^{(1)} + \frac{1}{6}\vec{\triangle} \Big( \ma{G} \cdot \vec{f}^{(1)} \Big) \:.
\end{equation}
This implies that the contribution $\vec{f}'^{(1)}=-\vec{f}^{(1)}$ to the force on the particle is the Stokes drag corresponding to the slip velocity $-\vec{\mathcal{U}}^{(1)}$, namely
\begin{equation}
\vec{f}'^{(1)}(t) =   6\pi \:\vec{\mathcal{U}}^{(1)}(t) = -6 \pi \int_0^t \ma{K}(t-\tau)  \cdot  \frac{\mbox{d} \vec{f}^{(0)}}{\mbox{d}\tau} \mbox{d}\tau\,,
\label{res_F1_linear}
\end{equation}
where the second equality stems from (\ref{T1hat}). Since $\vec{f}'^{(1)}$ is generally not collinear to $\vec{f}'^{(0)}$, the possibly nonzero component $\vec{f}'^{(1)}-\frac{\vec{f}'^{(1)}\cdot\vec{f}'^{(0)}}{\vec{f}'^{(0)}\cdot\vec{f}'^{(0)}}\vec{f}'^{(0)}$ represents a lift force. In the case of a sphere, and more generally of a body with a fore-aft symmetry, this is even the leading-order lift contribution, since symmetry considerations indicate that  no lift force can exist at $\mathcal{O}(\epsilon^0)$ for such bodies \citep{Bretherton1962}. \\
\indent For an arbitrarily shaped particle, the $\mathcal{O}(\epsilon)$-torque comprises a contribution proportional to the right-hand side of (\ref{res_F1_linear}). However, $\vec{\mathcal{U}}^{(1)}$ is a uniform velocity field. Therefore, just as $\vec{u}_s$ cannot induce any torque on a spherical particle in the creeping-flow limit, no $\mathcal{O}(\epsilon)$-torque may result from $\vec{\mathcal{U}}^{(1)}$, implying 
\begin{equation}
\vec{\tau}'^{(1)}(t) =  \bf{0}\,.
\label{res_tau1_linear}
\end{equation}
Obviously, this conclusion would not hold if the shape of the particle were such that its translational and rotational dynamics are coupled in the creeping-flow limit.\vspace{2mm}\\
\indent The $\mathcal{O}(\epsilon^2)$-problem is more complicated, owing to the $\mathcal{O}(\epsilon^2)$-terms on the left-hand side of (\ref{eqs_final}). More specifically, one has to solve
\begin{equation}
\boldsymbol{\nabla} \cdot  {\vec{w}}_{\mbox{\scriptsize in}}^{(2)}= \vec{0}\:,
\end{equation}
\begin{equation}
\frac{ \partial \vec{w}{\mbox{\scriptsize in}}^{(0)}}{\partial t} 
+ \ma{A}\cdot \vec{w}_{\mbox{\scriptsize in}}^{(0)} 
+ (\ma{A}\cdot \vec{r}) \cdot \boldsymbol{\nabla} \vec{w}_{\mbox{\scriptsize in}}^{(0)}
- \vec{u}_s \cdot \boldsymbol{\nabla} \vec{w}_{\mbox{\scriptsize in}}^{(0)} +  \vec{w}_{\mbox{\scriptsize in}}^{(0)} \cdot \boldsymbol{\nabla} \vec{w}_{\mbox{\scriptsize in}}^{(0)}   =   - \boldsymbol{\nabla} {p}_{\mbox{\scriptsize in}}^{(2)}  + \boldsymbol{\triangle}  
  {\vec{w}_{\mbox{\scriptsize in}}^{(2)}}\,,
 \label{eq_Linear_order2}
\end{equation}
together with the boundary conditions
\begin{equation}
{\vec{w}}_{\mbox{\scriptsize in}}^{(2)} =   \vec{0}  \quad \mbox{at}\quad r=1\,, 
\quad {\vec{w}}_{\mbox{\scriptsize in}}^{(2)} \sim  \vec{\mathcal{U}}^{(2)}(t)+\vec{\mathcal{T}}^{(2)}_{\rm{reg}} \quad \mbox{for} \quad r \to \infty\:. 
\label{BC:w_h}
\end{equation}
The above problem  is inhomogeneous. 
We seek its solution in the form a  particular (forced) solution, to which we add the uniform contribution $\vec{\mathcal{U}}_2(t)$, plus a solution of the homogeneous problem. In other words we set
\begin{equation}
 {p}_{\mbox{\scriptsize in}}^{(2)}  = p_p + p_h \,, \quad {\vec{w}_{\mbox{\scriptsize in}}^{(2)}} = \vec{w}_{p}+ \vec{\mathcal{U}}_2(t)  + \vec{w}_{h} \:.
 \label{ph}
\end{equation}
We obtain the formal solutions corresponding to $\vec{w}_{p}$ and $\vec{w}_{h}$ using Maple$^{\textregistered}$. Here we just outline the main steps of the procedure
but do not provide the final expressions since they are extremely lengthy. {\color{black}{Nevertheless the corresponding routines may be obtained on request from the authors. Moreover, a pivotal technical step in the building of these inner solutions turned out to be the generic determination of the Fourier transforms of functions involving negative or positive powers of $r$. Since this aspect may be of interest to some readers, we provide the corresponding results in supplementary material available at...}} \\
\indent To determine the particular solution $\vec{w}_{p}$, 
we first compute the Fourier transform ($\vec{\mathcal{F}}$) of the associated pressure in the form
\begin{equation}
\hat{p}_p = \frac{\mathtt{i} }{k^2} \left(\vec{k}\cdot \vec{\mathcal{F}}\left(\frac{ \partial \vec{w}{\mbox{\scriptsize in}}^{(0)}}{\partial t} 
+ \ma{A}\cdot \vec{w}_{\mbox{\scriptsize in}}^{(0)} 
+ (\ma{A}\cdot \vec{r}) \cdot \boldsymbol{\nabla} \vec{w}_{\mbox{\scriptsize in}}^{(0)}
- \vec{u}_s \cdot \boldsymbol{\nabla} \vec{w}_{\mbox{\scriptsize in}}^{(0)} +  \vec{w}_{\mbox{\scriptsize in}}^{(0)} \cdot \boldsymbol{\nabla} \vec{w}_{\mbox{\scriptsize in}}^{(0)} \right)\right) \,,
\end{equation}
with $\mathtt{i}^2=-1$. This allows us to obtain the particular solution for the velocity in Fourier space as
\begin{equation}
\hat{\vec{w}}_p = - \hat{\ma{G}} \cdot \vec{\mathcal{F}}\left(\frac{ \partial \vec{w}{\mbox{\scriptsize in}}^{(0)}}{\partial t} 
+ \ma{A}\cdot \vec{w}_{\mbox{\scriptsize in}}^{(0)} 
+ (\ma{A}\cdot \vec{r}) \cdot \boldsymbol{\nabla} \vec{w}_{\mbox{\scriptsize in}}^{(0)}
- \vec{u}_s \cdot \boldsymbol{\nabla} \vec{w}_{\mbox{\scriptsize in}}^{(0)} +  \vec{w}_{\mbox{\scriptsize in}}^{(0)} \cdot \boldsymbol{\nabla} \vec{w}_{\mbox{\scriptsize in}}^{(0)} \right) \:.
\end{equation}
The inverse Fourier transform then yields the particular solution in the physical space as
\begin{equation}
\vec{w}_p = \vec{\mathcal{F}}^{-1} ( \hat{\vec{w}}_p) \quad \mbox{and} \quad p_p = \vec{\mathcal{F}}^{-1} ( \hat{p}_p)\:.
\label{back}
\end{equation}
{\color{black}{Note that, in line with the comments made in \S\,\ref{outerp}, the contribution of the Oseen-like terms resulting from the leading-order solution $\vec{w}_{\mbox{\scriptsize in}}^{(0)}$ is accounted for in (\ref{eq_Linear_order2})-(\ref{back}).}}
Next, we consider the homogeneous solution $\ve w_h$. It satisfies
\begin{equation}
\boldsymbol{\nabla} \cdot \vec{w}_{h} = 0\:,\quad  - \boldsymbol{\nabla} {p}_h   + \boldsymbol{\triangle}  
  \vec{w}_{h} = \vec{0}\:, 
\end{equation}
together with the boundary conditions
\begin{equation}
\label{eq:wh}
\vec{w}_{h}  = - \vec{w}_p - \vec{\mathcal{U}}_2(t) \quad\mbox{at}\quad r=1\,,\quad \vec{w}_{h} \sim \vec{0} \quad\mbox{for} \quad r \to \infty\,.
\end{equation}
As can be seen, the particular solution computed previously now appears as a  boundary condition on the particle surface,
together with the uniform velocity correction $\ve{\mathcal{U}}_2(t)$ resulting from the outer solution. We finally obtain the solution of (\ref{eq:wh}) using Lamb's expansion \citep[][art. 336]{Lamb1932}. {\color{black}{Note that, unlike the calculation of $\vec{w}_{\mbox{\scriptsize out}}$, the evaluation of the inner solution is left unchanged if $\ma{A}$ is time-dependent. Therefore, the second-order inner contributions to the force and torque discussed below are valid even if the underlying linear flow is not stationary.}}

\begin{table}
\caption{Summary of all contributions
to the second-order force and torque, the former expressed with respect to $\vec{f}^{(0)} = 6\pi\vec{u}_s$. Terms arising from the regular and singular parts of the solution are given in the upper and lower tables, respectively. 
We also distinguish between the contributions to the force induced by the viscous part of the stress tensor (e.g. $\vec{f}_\nu$)  and 
those induced by the pressure (e.g. $\vec{f}_p$).  
The total second-order hydrodynamic force is the sum of these two contributions.
}
\mbox{}\\[1cm]
\centering {\bf Regular terms}
$$
\begin{array}{c | c | c | c | c | c }
\hline 
 & 
  & & &\\
 & \partial_t \vec{w}_{\mbox{\scriptsize in}}^{(0)} & \ma{A}\cdot\vec{w}_{\mbox{\scriptsize in}}^{(0)} + (\ma{A}\cdot \vec{r}) \cdot \boldsymbol{\nabla} \vec{w}_{\mbox{\scriptsize in}}^{(0)} & 
 -\vec{u}_s \cdot \boldsymbol{\nabla} \vec{w}_{\mbox{\scriptsize in}}^{(0)}  &  \vec{w}_{\mbox{\scriptsize in}}^{(0)} \cdot \boldsymbol{\nabla}\vec{w}_{\mbox{\scriptsize in}}^{(0)} \\
 &  
  & & &\\
\hline 
\hline
 & 
  & & &\\
\vec{f}'^{(2)}_\nu 
& \frac{2}{3} \frac{\text{d} \vec{f}^{(0)}}{\text{d}t} & \frac{2}{5} \ma{S} \cdot \vec{f}^{(0)} & \vec{0} & -\frac{1}{15} \ma{S} \cdot \vec{f}^{(0)} + \frac{1}{12} \boldsymbol{\omega}_s \times \vec{f}^{(0)} \\
 &  
 & & &\\
\hline 
 & 
  & & &\\
 \vec{f}'^{(2)}_p  
 & \frac{2}{9} \frac{\text{d} \vec{f}^{(0)}}{\text{d}t} & \frac{13}{45} \ma{S} \cdot \vec{f}^{(0)}+\frac{2}{9} \boldsymbol{\Omega}\times \vec{f}^{(0)} &  -\frac{1}{9}  \ma{S} \cdot \vec{f}^{(0)} + \frac{1}{9} \boldsymbol{\omega}_s \times \vec{f}^{(0)} & \frac{1}{360}\ma{S} \cdot \vec{f}^{(0)}-\frac{1}{36} \boldsymbol{\omega}_s\times \vec{f}^{(0)} \\
&  
& & &\\
\hline 
& 
& & &\\
 \vec{f}'^{(2)}
 & \frac{8}{9} \frac{\text{d} \vec{f}^{(0)}}{\text{d}t} & \frac{31}{45} \ma{S} \cdot \vec{f}^{(0)}+\frac{2}{9} \boldsymbol{\Omega}\times \vec{f}^{(0)} &  -\frac{1}{9}  \ma{S} \cdot \vec{f}^{(0)} + \frac{1}{9} \boldsymbol{\omega}_s \times \vec{f}^{(0)} & -\frac{23}{360}\ma{S} \cdot \vec{f}^{(0)}+\frac{1}{18} \boldsymbol{\omega}_s\times \vec{f}^{(0)} \\
&   
 & & &\\
\hline 
\hline
& 
 & & &\\
 \vec{\tau}'^{(2)}
 & -\frac{8\pi}{3}\frac{\text{d}\boldsymbol{\omega}_s}{\text{d}t}  & -\frac{8\pi}{3} \ma{S} \cdot \boldsymbol{\Omega} +\frac{16\pi}{15} \ma{S}\cdot  \boldsymbol{\omega}_s &  -\frac{1}{3}  \vec{u}_s \times \vec{f}^{(0)} = \vec{0} & -\frac{\pi}{15}\ma{S} \cdot \boldsymbol{\omega}_s \\
 &  
  & & &\\
   \hline
\end{array}
$$\mbox{}\\[2cm]
\centering {\bf Singular terms,} with\, 
$ \vec{\mathcal{U}}_2(t) =  - \int_0^{t} \ma{K}(t-\tau) \cdot  \frac{\mbox{d}}{\mbox{d}\tau}\left( \int_0^{\tau} 6\pi \:\ma{K}(\tau-\tau') \cdot \frac{\mbox{d}\vec{f}^{(0)}(\tau')}{\mbox{d}\tau'} \mbox{d}\tau' \right) \mbox{d} \tau $
$$
\begin{array}{c | c | c | c | }
\hline 
 & & & \\
& 
 & \mbox{Short-time behaviour: } \displaystyle{\ma{K}(t) \sim \frac{1}{6\pi}\frac{\ma{I}}{\sqrt{\pi t}}} &\mbox{Long-time behaviour: } \displaystyle{\ma{K}(t) \to \overline{\ma{K}}}   \\  
  & & &\\
\hline 
\hline 
  & & &\\
\vec{f}'^{(2)}_\nu & 4\pi \vec{\mathcal{U}}_2(t) 
& -\frac{2}{3}\frac{\text{d} \vec{f}^{(0)}}{\text{d}t}  & - 4\pi\:\overline{\ma{K}}\cdot (6\pi \overline{\ma{K}})\cdot \vec{f}^{(0)} \\
 & & &\\
\hline 
  & & &\\
 \vec{f}'^{(2)}_p  
 & 2\pi  \vec{\mathcal{U}}_2(t)  
 & - \frac{1}{3} \frac{\text{d} \vec{f}^{(0)}}{\text{d}t} & - 2\pi\:\overline{\ma{K}}\cdot (6\pi \overline{\ma{K}})\cdot \vec{f}^{(0)} \\ 
& & &\\
\hline 
& & &\\
 \vec{f}'^{(2)}
& 6\pi \vec{\mathcal{U}}_2(t) 
& -\frac{\text{d} \vec{f}^{(0)}}{\text{d}t} & - 6\pi\:\overline{\ma{K}}\cdot (6\pi \overline{\ma{K}})\cdot \vec{f}^{(0)} \\
 & & &\\
\hline 
\hline 
\end{array}
$$
\label{Tab1}
\end{table}

\subsection{Second-order force and torque}
Using the successive solutions described above, the second-order contributions
to the force and torque acting on the particle may be computed from the standard definitions
\begin{equation}
{\vec{f}'}^{(2)} = 
\int_\mathcal{S} \bbsigma^{(2)}_{\mathrm{in}} \cdot \vec{n} \:\mbox{d} \mathcal{S}\,,
\quad {\boldsymbol{\tau}'}^{(2)} = 
\int_\mathcal{S} \vec{r} \times \bbsigma^{(2)}_{\mathrm{in}} \cdot \vec{n} \:\mbox{d} \mathcal{S}\,.
\quad 
\end{equation}
Here $\vec{n}$ is the outward unit normal to the particle, and the second-order stress tensor $\bbsigma^{(2)}_{\mathrm{in}}$ is defined as
$
\bbsigma^{(2)}_{\mathrm{in}} = - p_{\mathrm{in}}^{(2)} \ma{1} + 2 \ma{S}_{\mathrm{in}}^{(2)}
$, where $\ma{S}_{\mathrm{in}}^{(2)}$ is the symmetric part of the velocity-gradient tensor based on the velocity field $\vec{w}_{\mathrm{in}}^{(2)}$. 
The final result for the second-order force and  torque resulting from the inner solution is found to be
 \begin{equation}
\begin{split}
\vec{f}'^{(2)}(t) = &  \frac{16\pi}{3} \frac{\mbox{d}\vec{u}_s}{\mbox{d}t} +
  \frac{37\pi}{12} \ma{S} \cdot \vec{u}_s  + \frac{\pi}{3} \:\left(4\boldsymbol{\Omega}+3\boldsymbol{\omega_s} \right)\times \vec{u}_s
  + 6\pi \vec{\mathcal{U}}_2(t) \,
\end{split}
\label{Force}
 \end{equation}
 \begin{equation}
 \boldsymbol{\tau}'^{(2)}(t) = -\frac{8\pi}{3}\frac{\mbox{d}\boldsymbol{\omega}_s }{\mbox{d}t} -\frac{8\pi}{3}
  \ma{S}\cdot \boldsymbol{\Omega}  + \pi \:\ma{S}\cdot \boldsymbol{\omega}_s\: 
  \label{Torque}
 \end{equation}
 Equations (\ref{Force}) and (\ref{Torque}) are the main results of the paper.  {\color{black}{For the reason mentioned above,}} the expression for the torque and the inner contribution to the force are valid for an arbitrary linear background flow, i.e. an arbitrary combination of a uniform straining motion and a solid-body rotation, {\color{black}{both of which may possibly be time-dependent. In contrast, for the reason explained in \S\,\ref{outerp}, we were only able to obtain the far-field uniform corrections $\vec{\mathcal{U}}_1(t)$ and $\vec{\mathcal{U}}_2(t)$ in the presence of a steady linear component of the carrying flow, which restricts the general result for the force on the particle to this subclass of flows. }}
 It must also be stressed that, because of the nonlinearity of the outer problem, the kernel $\ma{K}$ must be computed separately for each given linear flow. This kernel is already known explicitly for pure shear, solid-body rotation and planar elongation (see CMM). \\
 \indent The quadratic contributions in (\ref{Force}) and (\ref{Torque}) involve all possible combinations of $\ma{S}$, $\boldsymbol\Omega$, $\omega_s$ and $\vec{u}_s$ allowed by symmetry constraints, with the exception of $\boldsymbol\Omega\times\boldsymbol{\omega}_s=\boldsymbol\Omega\times\boldsymbol{\omega}_p$ in (\ref{Torque}). Note that $\vec{u}_s$ does not appear in (\ref{Torque}). This is because the torque is an axial (or pseudo-) vector while $\vec{u}_s$ is a polar (or true) vector, and no axial vector can be formed by combining quadratically $\vec{u}_s$ with $\boldsymbol\Omega$, $\boldsymbol{\omega}_s$ or $\ma{S}$. Note also that $\vec{\mathcal{U}}_2$ does not contribute to the second-order torque, for reasons identical to those already discussed in connection with $ \boldsymbol{\tau}'^{(1)}$. The presence of the term $ \pi \boldsymbol{\omega}_s\times\vec{u}_s$ in (\ref{Force}) is noticeable. This contribution, which represents a lift force, is the extension to linear flows of that obtained by \cite{rubinow1961transverse} for a sphere rotating and translating in a fluid at rest. This force may be thought of as the visco-inertial analogue of the inviscid Magnus lift force. It results from the coupling of the translational and rotational velocity dynamics operated in the hydrodynamic force by advective effects, a feature that does not exist in the creeping-flow regime, owing to the geometrical symmetries of the particle. \\
 \indent In \S\,\ref{sec:sol}, we showed that the $\mathcal{O}(\epsilon^2)$-problem is inhomogeneous, i.e. non-zero terms arise on the left-hand side of (\ref{eq_Linear_order2}). 
The solution of the problem is linear with respect to these terms.  In order to determine this full solution, one can therefore solve a succession of `elementary' problems, considering first for instance only the unsteady term
$\partial_t \vec{w}^{(0)}_{\mathrm{in}}$, then only terms $\ma{A}\cdot\vec{w}_{\mbox{\scriptsize in}}^{(0)} + (\ma{A}\cdot \vec{r}) \cdot \boldsymbol{\nabla} \vec{w}_{\mbox{\scriptsize in}}^{(0)}$, and so on.  The full solution is obtained  by summing up these partial solutions.  With this procedure, one can trace back which contribution to the force is due to the pressure gradient, which is of viscous origin, etc.
Only viscous stresses contribute to the torque since $\vec{r}$ and $\vec{n}$ are collinear for a sphere. 
The origin of the various contributions to the second-order force and torque is summarized in table \ref{Tab1}.

\section{Discussion}
\label{disc}
\subsection{The $\mathcal{O}(\epsilon^2)$-torque}
At leading order, the angular velocity of a torque-free spherical particle immersed in a linear flow 
 is dictated by the vorticity of the undisturbed carrying flow. Indeed, for such a particle (\ref{eq:t0}) and (\ref{res_tau1_linear}) imply
$\boldsymbol{\omega}_p = \boldsymbol{\Omega} + \mathcal{O}(\epsilon^2)$. Now, considering the long-time limit of (\ref{Torque}), the $\mathcal{O}(\epsilon^2)$-disturbance-induced torque on such a particle is 
\begin{equation}
\label{eq:so}
\boldsymbol{\tau}'^{(2)} = -\frac{8\pi}{3}
  \ma{S}\cdot \boldsymbol{\Omega} \:. 
\end{equation}
This second-order torque is nonzero only if the base flow is three-dimensional {\color{black}{and has a nonzero vorticity (which implies that it is unsteady for the reason discussed at the beginning of \S\,\ref{problem})}}, since $ \ma{S}\cdot \boldsymbol{\Omega}$ is a vortex-stretching term vanishing in a two-dimensional flow. Note that computing the spatial average of $\vec{r}\times\frac{\mbox{D} \vec{U}^\infty}{\text{D} t}$ over the particle volume reveals that the base flow generally brings a complementary contribution to the second-order torque, namely
\begin{equation}
\label{eq:soinf}
\boldsymbol{\tau}_b= -\frac{8\pi}{15}\epsilon^2\,
  \ma{S}\cdot \boldsymbol{\Omega} \:. 
\end{equation}
 Adding (\ref{eq:so}) and (\ref{eq:soinf}), one concludes that the total second-order torque resulting from the velocity gradients of a general three-dimensional linear flow is  $-\frac{16\pi}{5}
  \ma{S}\cdot \boldsymbol{\Omega}$, as derived by \cite{candelier2016angular} for a neutrally buoyant particle. Since the translational dynamics does not influence the torque for the symmetry reasons discussed above, this result is unchanged if the particle is not neutrally buoyant, as (\ref{Torque}) indicates. Finally, setting the total torque $\boldsymbol{\tau}'^{(0)}+\epsilon\boldsymbol{\tau}'^{(1)}+\epsilon^2\boldsymbol{\tau}'^{(2)}+\boldsymbol{\tau}_b$ to zero yields the angular velocity of a torque-free particle as
\begin{equation}
\label{eq:taut}
\boldsymbol{\omega}_p \approx \boldsymbol{\Omega}  - \frac{2}{5}\epsilon^2  \ma{S}\cdot \boldsymbol{\Omega} \,.
 \end{equation}

\vspace{5pt} 
No history (or memory) contribution appears in the expression of the second-order torque provided by (\ref{Torque}). This is surprising at first glance, since the problem reduces to the unsteady Stokes equation when unsteady effects are large enough for advective terms to become negligible near the particle.
In this limit, 
  \cite{Feuillebois1978} (see also \cite{Gatignol1983}) computed the torque acting on a spinning sphere, showing that
\begin{equation}
\boldsymbol{\tau}'(t)= -8 \pi \boldsymbol{\omega}_s(t) - \frac{8\pi}{3} \int_0^t \frac{\mbox{d} \boldsymbol{\omega}_s}{\mbox{d}\tau} \Bigg[ \frac{1}{\sqrt{\frac{\pi (t-\tau)}{\epsilon^2}}} - \exp\left(\frac{t-\tau}{\epsilon^2}\right) \mbox{erfc}\Bigg(\sqrt{\frac{t-\tau}{\epsilon^2}}\,\,\Bigg) \Bigg]\mbox{d}\tau\:.
\label{torque_Gat}
\end{equation}
The first term on the right-hand side corresponds to the leading-order torque (\ref{eq:t0}). 
To compare the second term in (\ref{torque_Gat}) with the prediction (\ref{Torque}), it must be borne in mind that all terms on the left-hand side of (\ref{eq:outer_tp}) are assumed small compared to unity, so that the present theory is valid provided
$\rm{Sl} \ll \rm{Re}_s^{-1} = \epsilon^{-2}$.
We show in appendix \ref{bubble} that in the limit $\epsilon\rightarrow0$, the term in square brackets in (\ref{torque_Gat}) tends toward $\epsilon^2 \delta(t)$, 
 with $\delta(t)$ the delta function (see (\ref{asymdelta})).
 In the limit of small $\epsilon$, the memory term in (\ref{torque_Gat}) therefore tends towards the first term on the right-hand side of (\ref{Torque}), namely
 \begin{equation}
 - \frac{8\pi}{3} \int_0^t \frac{\mbox{d} \boldsymbol{\omega}_s}{\mbox{d}\tau} \Bigg[ \frac{1}{\sqrt{\frac{\pi (t-\tau)}{\epsilon^2}}} - \exp\left(\frac{t-\tau}{\epsilon^2}\right) \mbox{erfc}\left(\sqrt{\frac{t-\tau}{\epsilon^2}}\right) \Bigg]\mbox{d}\tau \to - \frac{8\pi}{3} \epsilon^2\frac{\mbox{d} \boldsymbol{\omega}_s}{\mbox{d}t}\:.
\end{equation}
  This shows that (\ref{Torque}) and (\ref{torque_Gat}) are consistent, and the first contribution on the right-hand side of the former is what is left of the history torque at $\mathcal{O}(\epsilon^2)$ when $\epsilon$ is small.
  In other words, the memory term derived by \cite{Feuillebois1978} converges --  after a short transient of the order of the viscous time
 scale -- toward the expression obtained here, in which only the instantaneous angular acceleration $\frac{\mbox{d} \boldsymbol{\omega}_s}{\mbox{d}t}$ appears. This observation also explains why the theory describing the $\mathcal{O}(\text{Re}_s)$-effects on the angular dynamics of
  particles immersed in a stationary shear flow \citep{einarsson2015rotation} does not contain a memory term. 

\subsection{Second-order force at short times}
\label{shortt}
We now examine the force acting on the particle at short time {\color{black}{with respect to the time scale of the background flow}}, i.e. non-dimensional times such that $\epsilon^2\ll t\ll1$. {\color{black}{Within this time interval, the vorticity generated at the particle surface at $t=0$ has already diffused several radii away from the particle but has not yet entered the wake located at distances $r\gtrsim\epsilon^{-1}$.}} 
At order $\epsilon$, it is known that unsteady inertial effects are responsible for the existence of a history force, the expression of which takes the form of a convolution product between a time-dependent kernel $\ma{K}(t)$ and the particle relative acceleration (or more precisely the force $\vec{f}'^{(0)}$ acting on the particle in the creeping-flow limit). As discussed in \S\,\ref{outerp}, a term with a similar structure exists at $\mathcal{O}(\epsilon^2)$, namely $\vec{\mathcal{U}}_2$, the kernel involved (twice) in the expression of this second-order term being the same as that involved in the $\mathcal{O}(\epsilon)$-memory term. Both kernels depend on the specific undisturbed flow under consideration. However, at short times {\color{black}{($\epsilon^2\ll t\ll1$)}} and for any linear flow, $\ma{K}$ is closely approximated by the Basset-Boussinesq kernel, namely 
 \begin{equation}
 6 \pi \ma{K}(t) \sim  \frac{\ma{I}}{\sqrt{\pi t}}\:.
\end{equation}
In this case, $\vec{\mathcal{U}}_2(t)$ simplifies to
\begin{equation}
\vec{\mathcal{U}}_2(t) = \int_0^t \frac{1}{6 \pi} \frac{1}{\sqrt{\pi (t-\tau)}} \frac{\mbox{d}}{\mbox{d}\tau} \left( \int_0^\tau  \frac{1}{\sqrt{\pi (\tau-\tau')}} \frac{\mbox{d} \vec{f}^{(0)}}{\mbox{d}\tau'} \right) \mbox{d}\tau=   \frac{1}{6 \pi}  \frac{\mbox{d} \vec{f}^{(0)}}{\mbox{d} t } =    \frac{\mbox{d} \vec{u}_s}{\mbox{d} t }\:.
\label{res_frac_der}
\end{equation}
This can be shown using the definition of the fractional derivative $\mbox{d}^{1/2}/\mbox{d} t^{1/2}$ or, equivalently, Laplace transform.  Remarkably, the double convolution product reduces to a 
 simple `local' term (with respect to time) expressible in the form of a  time derivative of the relative velocity.  Using (\ref{res_frac_der}), (\ref{Force})  simplifies in
 \begin{equation}
\vec{f}'^{(2)} = -  \frac{2\pi}{3} \frac{\mbox{d}\vec{u}_s}{\mbox{d}t} +
  \frac{37\pi}{12} \ma{S} \cdot \vec{u}_s  + \frac{\pi}{3} \:\left(4\boldsymbol{\Omega}+3\boldsymbol{\omega_s} \right)\times \vec{u}_s \,.
\label{Force2}
 \end{equation}
 The first term on the right-hand side corresponds to the added-mass force if the undisturbed flow is uniform, see (\ref{BBO}). All quadratic terms in (\ref{Force2}) come from the inner solution. In contrast, the added-mass term results mostly from $\vec{\mathcal{U}}_2(t)$, i.e. from the second-order outer solution, as its sign differs from that of the corresponding term provided by the inner solution in (\ref{Force}). {\color{black}{This situation contrasts with that found in (\ref{Torque}) for the torque, for which the contribution proportional to the time derivative $\frac{\mbox{\small{d}}\boldsymbol{\omega}_s}{\mbox{\small{d}}t}$ results entirely from the inner solution. Therefore, one has to conclude that the inner and outer regions of the disturbance may both contribute to the `remains' of history effects, the region providing the dominant contribution to the corresponding time derivative term differing according to the load component under consideration.}}\\
\indent Equation (\ref{Force2}) may be recast in the form
\begin{equation}
\vec{f}'^{(2)} = - \frac{2\pi}{3} \left( \frac{{\rm d} \vec{v}_p}{{\rm d}t} -  \frac{\mbox{D} \vec{U}^\infty}{\mbox{D} t} \Big|_{\vec{x}_p}\right)  
+ \frac{15\pi}{4} \ma{S} \cdot \vec{u}_s  + \pi  \left(  \boldsymbol{\Omega} + \boldsymbol{\omega}_p \right)\times\vec{u}_s \,,
\label{Force3}
 \end{equation} 
 \begin{equation}
 \equiv- \frac{1}{2} \left( \frac{{\rm d} \vec{v}_p}{{\rm d}t} -  \frac{\mbox{D} \vec{U}^\infty}{\mbox{D} t} \Big|_{\vec{x}_p}\right)  
+ \frac{45}{16} \ma{S} \cdot \vec{u}_s  -\frac{3}{4} \vec{u}_s\times\left( \nabla\times \vec{U}^\infty \right) \,,
\label{Force3}
 \end{equation} 
 
 using again the identity $\text{d}\vec{U}^\infty/\text{d}t\big|_{\vec{x}_p}=\text{D}\vec{U}^\infty/\text{D}t\big|_{\vec{x}_p}+\ma{A}\cdot \vec{u}_s$ implying  
$\mbox{d}\vec{u}_s/\mbox{d}t = {\rm d}\vec{v}_p/{\rm d}t  - \mbox{D}\vec{U}^\infty/\mbox{D}t\big|_{\vec{x}_p} - \ma{A}\cdot \vec{u}_s$. Equation (\ref{Force3}) for $\ve{f}'^{(2)}$ allows a direct comparison with the inviscid form (\ref{auton2}) of the disturbance-induced force. In (\ref{Force3}), the added-mass term (first term on the right-hand side) is identical to that known in the inviscid limit. 
 From (\ref{Force3}) and (\ref{auton2}), the difference $\vec{f}'^{(2)}-\vec{f}'_{inv} $ evaluates to
\begin{equation} 
\vec{f}'^{(2)}-\vec{f}'_{inv}=  \frac{15\pi}{4} \ma{S} \cdot \vec{u}_s + \pi\left(\boldsymbol{\Omega}-\boldsymbol{\nabla}\times   \vec{U}^\infty \right)\times \vec{u}_s  + \pi \boldsymbol{\omega}_p\times\vec{u}_s \:.
\label{eqdiff}
\end{equation}
Note that in (\ref{eqdiff}) the pre-factor of the inviscid lift force is $\pi$ instead of $\frac{2\pi}{3}$ in (\ref{auton2}). This is because we have to employ the short-time expression for this force to remain consistent with that considered for $\vec{f}'^{(2)}$. It has been shown  \citep{legendre1998lift} that in this limit, more precisely for $t\ll u_c/as\equiv\text{Re}_s/\text{Re}_p\sim1$, the lift coefficient (usually defined as $\frac{3}{4\pi}$ times the above pre-factor) is $\frac{3}{4}$ instead of $\frac{1}{2}$ in the steady state, which yields the above result. 
With this pre-factor for the inviscid lift force, the second term on the right-hand side of (\ref{eqdiff}) may be simplified as $\frac{\pi}{2} \vec{u}_s \times \left(\boldsymbol{\nabla}\times   \vec{U}^\infty \right)$ using the identity $\boldsymbol{\Omega}=\frac{1}{2}( \boldsymbol{\nabla} \times \vec{U}^\infty)$ but the original form appears more suitable for the discussion that follows. Not surprisingly, $\vec{f}'^{(2)}$ does not coincide with $\vec{f}'_{inv}$, although the two added-mass terms do. The last contribution on the right-hand side of (\ref{eqdiff}) gives an immediate clue to understand where the differences come from. Indeed, the particle rotation does not matter in the inviscid limit, since (i) rotation does not displace any fluid in the specific case of a sphere, and (ii) the fluid is free to slip at the particle surface. Therefore, no force proportional to $\boldsymbol{\omega}_p$ can exist in this limit, which is in stark contrast with the visco-inertial regime considered here, in which the no-slip condition forces the surrounding fluid to follow the particle rotation, yielding the nonzero Rubinow-Keller lift force. \\
\indent The no-slip condition is also responsible for the contribution $\frac{15\pi}{4} \ma{S} \cdot \vec{u}_s$ which has no counterpart in the inviscid limit, an indication that it results from the influence of the ambient strain on the disturbance generated in the sphere vicinity by the no-slip condition. Note that in general this term may contribute to both the drag and the lift components of the total force. The previous argument regarding the role of the no-slip condition holds for the contribution $-\pi\vec{u}_s \times \boldsymbol{\Omega}$ but the presence of the inviscid force $\pi\vec{u}_s \times (\boldsymbol{\nabla}\times   \vec{U}^\infty)$ makes the point a little bit more subtle. As is well known, the inviscid shear lift force results from the distortion of the vorticity $\boldsymbol{\nabla}\times  \vec{U}^\infty $ contained in the background flow as it is transported along the curved streamlines around the (inviscid) sphere. \cite{lighthill1956drift} gave an illuminating quantitative description of how this inviscid stretching/tilting mechanism results in the generation of a pair of counter-rotating streamwise vortices in the wake of a sphere immersed in a linear shear flow. This mechanism subsists qualitatively at finite Reynolds number. However, when the no-slip condition applies at the sphere surface, the local kinematic structure of the undisturbed flow, characterized by $ \ma{S}$ and $\boldsymbol{\Omega}$, influences the velocity disturbance required to satisfy this condition, hence the near-surface vorticity. The term $-\pi\vec{u}_s \times \boldsymbol{\Omega}$ in (\ref{eqdiff}), or at least a part of it, results from this effect. It may be that, owing to the identity $\boldsymbol{\Omega}=\frac{1}{2}( \boldsymbol{\nabla} \times \vec{U}^\infty)$, a nonzero part of this term rather results from what remains at low-but-finite Reynolds number of the aforementioned stretching/tilting mechanism of the upstream vorticity, and that the two combine to yield the pre-factor $-\pi$. In any case, there is no reason for the entire finite-Reynolds-number contribution to be identical to the inviscid lift force. Noting that the magnitude of the former is smaller than that of the latter ($-\frac{\pi}{2}$ instead of $-\pi$ once expressed with respect to $ \vec{u}_s \times (\boldsymbol{\nabla}\times \vec{U}^\infty)$), we may even conclude that the alteration of the near-surface disturbance by the rotational component $\boldsymbol{\Omega}$ of the undisturbed velocity gradient yields a contribution to the force whose sign is opposite to that of the inviscid lift force.

\subsection{Stationary limit: pure straining flows}
\label{statlim}
 In the stationary limit, the slip velocity between the particle and the fluid no longer varies. Similarly, $\tfrac{{\rm d}\vec{f}^{(0)} }{{\rm d}t}$ tends to zero and the singular term in (\ref{Force}) becomes 
$$
 6\pi \vec{\mathcal{U}}_2 \to  -(6\pi \overline{\ma{K}}) \cdot (6\pi \overline{\ma{K}}) \cdot \vec{f}^{(0)}\:,
$$
where $6\pi \overline{\ma{K}}$ may be thought of as the steady resistance tensor  induced by fluid inertia effects. The second-order force  resulting from the flow disturbance then becomes
\begin{equation}
\vec{f}'^{(2)} = \frac{37 \pi}{12} \ma{S} \cdot \vec{u}_s 
-  \frac{\pi}{6} \vec{u}_s \times \left( \boldsymbol{\nabla} \times \vec{U}^\infty\right) 
- \pi 
\vec{u}_s\times\boldsymbol{\omega}_p    - 6\pi  (6\pi \overline{\ma{K}}) \cdot (6\pi \overline{\ma{K}}) \cdot \vec{u}_s \:.
\label{stat}
\end{equation}

\noindent Again, this expression is general in that it is valid in any linear flow. The actual difficulty to use it in practice is that the explicit expression for the long time kernel is generally not known, except in some canonical configurations.\vspace{2mm}\\
\indent We first specialize the above result to the case of a planar extensional flow, defined as 
\begin{equation}
\ma{A} = \left(\begin{array}{ccc} 1 & 0 & 0\\
0 & -1 & 0 \\
0 & 0 & 0\\
\end{array}\right) \:.
\label{strain0}
\end{equation}
The corresponding steady-state kernel was computed by CMM. However, a technical difficulty (a non-convergence of a three-dimensional integral to be evaluated in Fourier space) prevented the computation of the component of the kernel corresponding to the compressional direction $\vec{e}_2$ beyond a certain time ($t\approx30$). The final result obtained in this reference reads
\begin{equation}
6\pi \overline{\ma{K}} \approx \left(\begin{array}{ccc} 
0.901 & 0 & 0\\
0 & [\overline{\ma{K}}]_2^2& 0 \\
0 & 0 & 0.420\\
\end{array}
\right)\:,
\label{elong}
\end{equation}
with $[\overline{\ma{K}}]_2^2$ presumably negative and larger than $1.48$ in absolute value. Fortunately, the diagonal nature of $\overline{\ma{K}}$ leaves the $\vec{e}_1$-component of the resulting force unaffected by this uncertainty, even at $\mathcal{O}(\epsilon^2)$. Let us assume that the particle is at rest at the location $\vec{x}_p=(1,x_{p2},x_{p3})$. This implies $\vec{u}_s=-\vec{e}_1+x_{p2}\vec{e}_2$, so that the first-order correction to the $\vec{e}_1$-component of the disturbance-induced force becomes $\ve f'^{(1)}\cdot\vec{e}_1=6\pi\vec{e}_1\cdot(6\pi\overline{\ma{K}})\cdot\vec{e}_1\approx16.98$. As (\ref{Torque}) shows, the undisturbed flow does not provide any contribution to the second-order torque because $\boldsymbol{\Omega}\equiv\bf{0}$. Therefore, a torque-free sphere does not rotate in the present flow whatever its transverse position $x_{p2}$, unless it has been given an initial nonzero rotation. If one sets $\boldsymbol{\omega}_p=\bf{0}$ in (\ref{Force}), the second-order correction to the $\vec{e}_1$-component of the disturbance-induced force is found to be 
\begin{equation}
\ve f'^{(2)}\cdot\vec{e}_1= \frac{37\pi}{12} \vec{e}_1\cdot(\ma{S} \cdot \vec{u}_s)   + 6\pi  \vec{e}_1\cdot(6\pi \overline{\ma{K}}) \cdot (6\pi \overline{\ma{K}})\cdot\vec{u}_s\approx5.615\,.
\label{stat1}
\end{equation}
\indent Another straining motion of interest is the uniaxial axisymmetric flow characterized by a constant stretching rate along the symmetry axis and a uniform compression in the plane perpendicular to it. This flow was not considered by CMM. The corresponding kernel is computed in appendix \ref{uniax}, for $\ma{S}=-(\vec{e}_1\vec{e}_1+\vec{e}_2\vec{e}_2)+2\vec{e}_3\vec{e}_3$. The steady-state result (\ref{kbar3d}) indicates that the axial component of the kernel is $ 6\pi[\overline{\ma{K}}]_3^3=6\pi\vec{e}_3\cdot\overline{\ma{K}}\cdot\vec{e}_3\approx1.26$. Therefore, considering a particle held fixed on the flow axis at the axial location $x_{p3}=\frac{1}{2}$ (which implies $\ve{u}_s=-\ve{e}_3$), the first-order correction to the force reads $\ve f'^{(1)}=  -6\pi  (6\pi \overline{\ma{K}}) \cdot \vec{u}_s\approx 23.75\, \vec{e}_3$, while the second-order correction is
\begin{equation}
\ve f'^{(2)}= \frac{37\pi}{12}\ma{S} \cdot \vec{u}_s -6\pi  (6\pi \overline{\ma{K}}) \cdot (6\pi \overline{\ma{K}}) \cdot \vec{u}_s\approx 10.55\, \vec{e}_3 \:.
\label{stat2}
\end{equation}
\indent In appendix \ref{bubble} we establish the counterpart of (\ref{stat2}) in the case of a spherical bubble at the surface of which the surrounding flow obeys a shear-free condition instead of the no-slip condition in (\ref{Eq1_bc}). The corresponding result reads 
\begin{equation}
\ve f'^{(2)}= \frac{4\pi}{3} \ma{S} \cdot \vec{u}_s -4\pi  (4\pi \overline{\ma{K}}) \cdot (4\pi \overline{\ma{K}}) \cdot \vec{u}_s\approx0.49\,\vec{e}_3\,.
\label{stat3}
\end{equation}
Comparing (\ref{stat2}) and (\ref{stat3}) reveals that the inner and outer contributions to $\ve f'^{(2)}$ both differ. In particular, the former is more than twice as small in the case of a bubble. This is a clear confirmation of the prominent role played by the vorticity produced at the particle surface in all contributions to the hydrodynamic force. \\
\indent The above comments call for a comparison between the predictions for $\ve f'^{(2)}$ and those for the inviscid force $\vec{f}'_{inv}$ given by (\ref{auton2}).  
 In a pure straining flow, $\vec{f}'_{inv}$ reduces to the added-mass force, i.e. $\vec{f}'_{inv}= \frac{2\pi}{3} \left(\frac{\mbox{D} \vec{U}^\infty}{\mbox{D} t} \Big|_{\vec{x}_p}-\frac{\mbox{d} \vec{v}_p}{\mbox{d}t}  \right)$. If the particle is held fixed, $\vec{U}^\infty\big|_{\vec{x}_p}=-\vec{u}_s$. Hence in the case of the planar flow (\ref{strain0}), setting again $\vec{u}_s\cdot\vec{e}_1=-1$, the $\vec{e}_1$-component of the inviscid force is $\vec{f}'_{inv}\cdot\vec{e}_1=\frac{2\pi}{3} \vec{e}_1\cdot\ma{S}\cdot\vec{e}_1=\frac{2\pi}{3}$. 
 Similarly, in the uniaxial straining flow, $\vec{f}'_{inv}\cdot\vec{e}_3=\frac{4\pi}{3}$. 
 Remarkably, although the inner contribution to $\ve f'^{(2)}$ and $\vec{f}'_{inv}$ are both directly proportional to $\ma{S}\cdot\ve{u}_s$, the former is negative for both types of straining flow and whatever the nature of the particle, while the latter is always positive. This observation sheds light on the debate summarized in \S\,\ref{intro} regarding the proper form of the added-mass force in the low-but-finite Reynolds number limit. Indeed, from a theoretical point of view, this turns out to be a `non-question' since the dynamic boundary condition at the particle surface, by nature a viscous effect, is found to produce contributions to the force proportional to $\ma{S}\cdot\ve{u}_s$, just as the added mass does if Taylor's expression involving the Lagrangian derivative of $\vec{U}^\infty$ holds. Having noticed this, our only hope to disentangle both effects stood in the splitting of the inner contributions to $\ve f'^{(2)}$ into viscous and pressure drag components in  (\ref{stat2}) and (\ref{stat3}), from which a comparison of all contributions for a solid sphere and a spherical bubble immersed in the same flow is possible. However, nothing emerged from this comparison, as the ratio of the viscous and pressure components (once the possible $\frac{4\pi}{3}$ added-mass contribution has been removed from the latter) differs for the two types of particle. Based on this unsuccessful attempt, we have to conclude that whether the added-mass force involves the fluid acceleration $\frac{\mbox{D} \vec{U}^\infty}{\mbox{D} t}\Big|_{\vec{x}_p} $ or the derivative of $\vec{U}^\infty$ following the particle motion, i.e. $\frac{\mbox{d} \vec{U}^\infty}{\mbox{d} t}\Big|_{\vec{x}_p} $, is undecidable at low-but-finite Reynolds number.\\
\indent Things are different at moderate-to-large Reynolds numbers, as the direct simulations of \cite{Magnaudet1995} showed. Their study considered the flow experienced by a rigid sphere or a spherical bubble held fixed in an axisymmetric uniaxial (or biaxial) straining flow over a wide range of Reynolds numbers. Whatever the Reynolds number $\text{Re}_p$ (which was varied from $0.05$ to $150$), the magnitude of the fluid acceleration and the nature of the particle, the results revealed the existence of two series of contributions dependent on both strain and slip effects. One, found in the pressure drag, has a magnitude independent of $\text{Re}_p$ and equal, to numerical accuracy, to that of the inviscid force predicted by (\ref{auton2}). In contrast, the other, found in both the pressure and viscous drag components,  depends on $\text{Re}_p$ in the same way as the corresponding drag component in the case the body is immersed in a uniform flow. Its origin was readily identified by considering the vorticity and pressure distributions at the surface of the particle. In particular, compared to the same particle in a uniform stream, the fluid acceleration was observed to decrease the surface vorticity on the front half and to increase it on the rear half. These findings establish that these contributions are a direct consequence of the changes in the vorticity distribution at the particle surface, which result from the strain component of the base flow. To summarize, these simulations made it clear that (i) added-mass effects due to the advective acceleration $\vec{U}^\infty\cdot\nabla\vec{U}^\infty\equiv -\ma{S}\cdot\vec{u}_s$ exist in the considered configuration whatever $\text{Re}_p$ and are unaffected by the Reynolds number; and (ii) the local strain characterizing the undisturbed flow modifies the near-surface vorticity distribution resulting from the dynamic boundary condition (no-slip or shear-free), which in turn yields additional contributions to the drag that depend linearly on $ \ma{S}$ but experience large variations with $\text{Re}_p$. Apart from their different behaviours with respect to $\text{Re}_p$, there is no way to separate the various $\ma{S}$-dependent contributions in the overall drag. The theoretical low-but-finite Reynolds number results obtained above in the planar and uniaxial straining flows 
are just another illustration of this coexistence of two categories of strain-dependent contributions to the drag resulting from two totally different physical mechanisms. Unlike the aforementioned simulations in which $\text{Re}_p$ was varied by several orders of magnitude, these contributions appear at the same $\mathcal{O}(\epsilon^2)$-order in the present asymptotic theory, and for this reason cannot be disentangled.
\subsection{Stationary limit: vortical flows}
\indent We now turn to the case of a linear shear flow in which the velocity gradient tensor $\ma{A} $ takes the form
\begin{equation}
\ma{A} = \left(\begin{array}{ccc} 0 & 0 & 1\\
0 & 0 & 0 \\
0 & 0 & 0\\
\end{array}\right)\,.
\end{equation}
\cite{saffman1965lift} computed the leading-order $\mathcal{O}(\epsilon)$-lift force in this flow for a sphere translating steadily along the direction of the undisturbed stream and possibly rotating perpendicular to it. 
He also computed the $\mathcal{O}(\epsilon^2)$-contributions to the lift force arising from the inner region of the disturbance. In contrast, he did not consider the $\mathcal{O}(\epsilon^2)$-force arising from the  singular term $\vec{\mathcal{U}}_2(t)$.  
Setting the slip velocity $ \vec{u}_s$ to $ -\vec{e}_1$ and the angular velocity of the particle to $\boldsymbol{\omega}_p= \omega_p \vec{e}_2$,  
Saffman's second-order result (his equation (2.17)) reads 
\begin{equation}
 {\vec{f}'}_S^{(2)} = \left( \pi \omega_p -\frac{11\pi}{8} \right)\vec{e}_3\:.
 \label{saffman2}
\end{equation}
The present theory allows us to compute all $\mathcal{O}(\epsilon^2)$-contributions, including those resulting from $\vec{\mathcal{U}}_2(t)$. Moreover, the orientation of the slip velocity may be kept arbitrary. Evaluating all terms in (\ref{Force}) leads to
\begin{equation}
\vec{f}'^{(2)} = \left(\begin{array}{ccc} 0 & 0 & \frac{41\pi}{24} + \pi \omega_p\\
0 & 0 & 0 \\
\frac{11 \pi}{8} - \pi \omega_p& 0 & 0\\
\end{array}\right) \cdot \vec{u}_s   - 6\pi  (6\pi \overline{\ma{K}}) \cdot (6\pi \overline{\ma{K}}) \cdot \vec{u}_s \:.
 \label{present2}
\end{equation}
 The stationary kernel $\overline{\ma{K}}$ is known to be (\citep{miyazaki1995drag}, CMM)
\begin{equation}
6\pi \overline{\ma{K}} \approx \left(\begin{array}{ccc} 
0.0737 & 0 & 0.9436\\
0 &0.5766 & 0 \\
0.3425 & 0 & 0.3269\\
\end{array}
\right)\:,
\label{skernel}
\end{equation}
so that the leading-order correction to the Stokes drag is $\vec{f}'^{(1)}=-6\pi(6\pi\overline{\ma{K}})\cdot\vec{u}_s$. In particular, the leading-order lift component $\vec{f}'^{(1)}\cdot\vec{e}_3$ in the configuration considered by Saffman is $\vec{f}'^{(1)}\cdot\vec{e}_3=6\pi(6\pi\vec{e}_3\cdot\overline{\ma{K}}\cdot\vec{e}_1)\approx6.456$.\\
\indent Using (\ref{skernel}), (\ref{present2}) provides the next-order force as
\begin{equation}
\vec{f}'^{(2)} = \left(\begin{array}{ccc} 0 & 0 & \frac{41\pi}{24} + \pi \omega_p\\
0 & 0 & 0 \\
\frac{11 \pi}{8} - \pi \omega_p& 0 & 0\\
\end{array}\right) \cdot \vec{u}_s   - \left(\begin{array}{ccc} 6.1942 & 0 & 7.1252\\
0 & 6.2669 & 0 \\
2.5862& 0 & 8.1062\\
\end{array}\right) \cdot \vec{u}_s\,\, \:.
 \label{present3}
\end{equation}
If the particle does not rotate, this reduces to
\begin{equation}
\vec{f}'^{(2)} = 
  - \left(\begin{array}{ccc} 6.1942 & 0 & 1.7583\\
0 & 6.2669 & 0 \\
-1.7335& 0 & 8.1062\\
\end{array}\right) \cdot \vec{u}_s \:.
 \label{present4}
\end{equation}
The off-diagonal components of $\ve f'^{(2)}$ yield the second-order lift force $\vec{f}'^{(2)}_{L}\approx-1.7583(\vec{u}_s\cdot\vec{e}_3)\vec{e}_1+1.7335(\vec{u}_s\cdot\vec{e}_1)\vec{e}_3$. In the specific case $\vec{u}_s=- \vec{e}_1$, (\ref{present4}) implies $\vec{f}'^{(2)}\cdot\vec{e}_3=-1.7335$, to be compared with $\vec{f}_S^{(2)}\cdot\vec{e}_3=-\frac{11}{8}\pi\approx-4.32$ according to (\ref{saffman2}). Hence, the actual magnitude of the second-order lift force is approximately $2.5$ times smaller than Saffman's incomplete prediction. It may be noticed that the two pre-factors involved in $\vec{f}'^{(2)}_{L}$ have very close values,  albeit with opposite signs, so that on a purely empirical basis the second-order lift force may be approximated as $\vec{f}'^{(2)}_{L}\approx1.746\,\vec{u}_s\times(\boldsymbol{\nabla} \times \vec{U}^\infty)$. This expression may be compared with the prediction for the inviscid disturbance-induced force  $\vec{f}'_{inv}$ in (\ref{auton2}), which in the present case reduces to the shear lift force $\vec{f}'_{inv}=-\frac{2\pi}{3}\vec{u}_s\times(\boldsymbol{\nabla} \times \vec{U}^\infty)$. Although the two expressions have the same structure and the two pre-factors have a comparable magnitude (since $\frac{2\pi}{3}\approx2.094$), they have opposite signs, a clear indication that the mechanism responsible for $\vec{f}'^{(2)}_{L}$ is totally different from the Lighthill mechanism discussed in \S\,\ref{shortt}.
\\
\indent Similarly, if the particle rotates freely, the torque-free condition implies $||\boldsymbol{\omega}_s|| =\mathcal{O}(\epsilon^2)$, i.e. $\boldsymbol{\omega}_p\approx\boldsymbol{\Omega}=\frac{1}{2}\vec{e}_2$, so that (\ref{present3}) simplifies to
\begin{equation}
\ve f'^{(2)} =   \left(\begin{array}{ccc} 
-6.1942 & 0 & -0.1876\\
0 & -6.2669 & 0 \\
0.1626 & 0 & -8.1061\\
\end{array}
\right)\cdot \vec{u}_s\:.
 \label{present5}
\end{equation}
Now the off-diagonal components of $\ve f'^{(2)}$ yield the second-order lift force $\vec{f}'^{(2)}_{L}\approx-0.1876(\vec{u}_s\cdot\vec{e}_3)\vec{e}_1+0.1626(\vec{u}_s\cdot\vec{e}_1)\vec{e}_3$. Again, the two pre-factors have very close values, so that this contribution may be empirically approximated in the form $\vec{f}'^{(2)}_{L}\approx0.175\vec{u}_s\times(\boldsymbol{\nabla} \times \vec{U}^\infty)$.
The pre-factors involved in $\vec{f}'^{(2)}_{L}$ are fairly small compared to those found in the non-rotating case because the contribution brought by the singular term $- 6\pi  (6\pi \overline{\ma{K}}) \cdot (6\pi \overline{\ma{K}})\cdot\vec{u}_s$ nearly balances the sum of the regular terms.  In particular, with $\vec{u}_s=- \vec{e}_1$, (\ref{present5}) implies $\vec{f}'^{(2)}_L=-0.1626\vec{e}_3$. In contrast, with the relevant angular velocity ${\omega}_p=\frac{1}{2}$, Saffman's partial result (\ref{saffman2}) yields $\vec{f}_S^{(2)}=-\frac{7}{8}\pi\vec{e}_3\approx-2.75\vec{e}_3$, which over-predicts the actual second-order lift force by more than one order of magnitude. The above two examples indicate that Saffman's incomplete result for the second-order correction to the lift force is grossly inaccurate. Therefore, practitioners using point-particle models incorporating the steady-state form of the Saffman lift force should either consider only the corresponding leading-order prediction or make use of (\ref{present5}) or (\ref{present4}) to estimate the second-order correction, depending on whether the particle rotates or not.\\

\indent Last, we turn to the solid-body rotation flow characterized by the velocity-gradient tensor
\begin{equation}
\ma{A} = \left(\begin{array}{ccc} 0 & -1 & 0\\
1 & 0 & 0 \\
0 & 0 & 0\\
\end{array}\right) \:.
\label{sbr}
\end{equation}
Since $\ma{S}\equiv\bf{0}$, there is no source term in (\ref{Torque}), so that the torque-free condition implies $\boldsymbol{\omega}_s=\bf{0}$ if the particle is not given an initial differential rotation. Therefore, once the slip velocity no longer varies, (\ref{Force}) reduces to
 \begin{equation}
\vec{f}'^{(2)} =  \frac{4\pi}{3} \boldsymbol{\Omega}\times \vec{u}_s- 6\pi  (6\pi \overline{\ma{K}}) \cdot (6\pi \overline{\ma{K}}) \cdot \vec{u}_s  \,.
\label{Forcer}
 \end{equation}
 For this flow, the steady-state kernel may be obtained in closed form and reads \citep[][CMM]{Gotoh90,Miyazaki95b} 
 \begin{equation}
6\pi \overline{\ma{K}} = \left(\begin{array}{ccc} 
\frac{3\sqrt{2}(19+9\sqrt{3})}{280}& -\frac{3\sqrt{2}(19-9\sqrt{3})}{280} & 0\\
\frac{3\sqrt{2}(19-9\sqrt{3})}{280} & \frac{3\sqrt{2}(19+9\sqrt{3})}{280}& 0 \\
0 & 0 & \frac{4}{7}\\
\end{array}
\right)\,.
\label{rotat}
\end{equation}
The off-diagonal terms of (\ref{rotat}) yield the leading-order rotational lift force $\vec{f}'^{(1)}_{L}=\frac{9\pi \sqrt{2}(19-9\sqrt{3})}{140}[(\vec{u}_s\cdot\vec{e}_2)\vec{e}_1-(\vec{u}_s\cdot\vec{e}_1)\vec{e}_2]$ $\approx0.980[(\vec{u}_s\cdot\vec{e}_2)\vec{e}_1-(\vec{u}_s\cdot\vec{e}_1)\vec{e}_2]$. Making use of (\ref{Forcer}), the second-order disturbance-induced force is found to be
\begin{equation}
 \vec{f}'^{(2)} =\frac{4\pi}{3} \vec{e}_3\times \vec{u}_s-6\pi  \left(\begin{array}{ccc} 
\frac{1539\sqrt{3}}{9800}& -\frac{531}{9800} & 0\\
\frac{531}{9800} & \frac{1539\sqrt{3}}{9800}& 0 \\
0 & 0 & \frac{16}{49}\\
\end{array}
\right)\cdot\vec{u}_s
\approx
\left(\begin{array}{ccc} 
-5.127&-3.167&0\\
3.167&-5.127&0\\
0&0&-6.155\\
\end{array}
\right)
\cdot\vec{u}_s\,.
\label{Forcerr}
\end{equation}
In particular, the second-order rotational lift force resulting from the off-diagonal terms of (\ref{Forcerr}) is $\vec{f}'^{(2)}_{L}\approx-3.167\,[(\vec{u}_s\cdot\vec{e}_2)\vec{e}_1-(\vec{u}_s\cdot\vec{e}_1)\vec{e}_2]\approx-1.584\,\vec{u}_s\times\left(\vec{\nabla} \times  \vec{U}^\infty\right)$, since $\vec{\nabla} \times  \vec{U}^\infty=2\vec{e}_3$ according to the definition (\ref{sbr}) of the velocity-gradient tensor.\\
\indent The pre-factor involved in $\vec{f}'^{(2)}_{L}$ is significantly larger than that found in $\vec{f}'^{(1)}_{L}$ and of opposite sign. The total lift force at the present order of approximation being $\vec{f}'_{L} = \epsilon\vec{f}'^{(1)}_{L} +\epsilon^2\vec{f}'^{(2)}_{L}$,  this force changes sign and becomes dominated by the second-order contribution beyond $\epsilon\approx0.31$, i.e. beyond a critical shear Reynolds number $\text{Re}_s\approx0.096$. Hence, under many practical conditions, one may expect a small particle held fixed in a solid-body rotation flow to experience a lift force of opposite sign to the one predicted by the leading-order estimate $\vec{f}'^{(1)}_{L}$. This surprising feature contrasts with what happens in the pure shear configuration. Indeed, (\ref{present5}) indicates that the total lift force experienced by a torque-free particle with a slip velocity $\vec{u}_s=-\vec{e}_1$ is in that case $\vec{f}'_{L} =(6.456\epsilon-0.1626\epsilon^2)\,\vec{e}_3$. Again, the first- and second-order contributions have opposite signs but the pre-factor of the $\mathcal{O}(\epsilon^2)$-term is much smaller than that of the $\mathcal{O}(\epsilon)$-term, so that $\vec{f}'_{L} $ never changes sign within the domain of validity of the present theory.\\
 \indent We may again compare $\vec{f}'^{(2)}_{L}$ with the inviscid force predicted by (\ref{auton2}), noting that in the present case $\frac{\mbox{D} \vec{U}^\infty}{\mbox{D} t}\Big|_{\vec{x}_p}\equiv\frac{1}{2}\vec{u}_s \times \left(\vec{\nabla} \times  \vec{U}^\infty\right)$ if the particle is at rest. For this reason, the added-mass and shear-induced lift forces in (\ref{auton2}) combine in the form of an `extended' lift force $\vec{f}'_{inv}=-\frac{\pi}{3}\: \vec{u}_s \times \left(\vec{\nabla} \times  \vec{U}^\infty\right)$. 
 This inviscid force and the second-order lift force $\vec{f}'^{(2)}_{L} $ are seen to have the same sign, which contrasts with what was noticed above in the pure shear configuration. However the magnitude of $\vec{f}'^{(2)}_{L} $ ($3.167$) is larger than that of $\vec{f}'_{inv}$ ($\approx2.094$), which once again underlines the role of the no-slip condition in the structure of the disturbance flow. That $\vec{f}'^{(2)}_{L} $ and $\vec{f}'_{inv}$ keep the same sign in a solid-body rotation flow but have opposite signs in a linear shear flow emphasizes the crucial importance of the alteration introduced in the near-surface disturbance by the presence of a nonzero strain in the latter case. More specifically, this change of sign suggests that the surface vorticity associated with this strain-induced disturbance contributes more to the lift force than any other mechanism affecting the vorticity distribution around the particle, especially those resulting from the stretching and tilting of the upstream vorticity $\vec{\nabla} \times  \vec{U}^\infty$.
\section{Summary and concluding remarks}
\label{conclu}
In this work, we computed the second-order inertial corrections to the creeping-flow approximation of the time-dependent hydrodynamic force and torque acting on a small rigid sphere immersed in a general steady linear flow. Our primary motivation was to obtain a consistent approximation for the force and torque in which all fundamental physical effects resulting from fluid inertia, such as shear- and spin-induced lift and added mass, are accounted for to $\mathcal{O}(\epsilon^2)$. To achieve this goal, we used and extended the methodology developed by CMM which employs asymptotic matching expansions and formulates the problem in a coordinate system co-moving with the carrying flow. Some of the second-order corrections derived here were already known, but others were not. Computing these corrections in a systematic fashion is much more difficult than obtaining the first-order corrections. This is on the one hand  
because the inner problem at $\mathcal{O}(\epsilon^2)$ is inhomogeneous, and on the other hand because the second-order singular contribution brought by the outer solution results from a double convolution product.\\ 
\indent Our results for the second-order force and torque are summarized in (\ref{Force})-(\ref{Torque}). However, it is only after the singular outer velocity correction $\vec{\mathcal{U}}_2(t)$ has been explicitly computed that one can fully appreciate how the different contributions combine in the second-order force, and of which physical effect they account for. This is especially true regarding terms involving the relative acceleration between the particle and fluid. In particular, the classical added-mass force in a uniform flow is found to result from the sum of an inner and an outer contribution evaluated in the short-time limit $t\ll1$, and it is dominated by the latter. That the outer contribution dominates indicates that the majority of the fluid instantaneously displaced when the particle or the fluid accelerates is located far from the body, i.e. at distances larger than the Saffman length. The expression (\ref{Torque}) for the torque also comprises a contribution involving the instantaneous relative angular acceleration, as if there were a nonzero rotational added-mass effect. This is of course not the case given the point-symmetric geometry of the considered particle, and this contribution is just due to the `remains' of the exact history torque beyond the initial stage that extends over a few viscous time units. Indeed, it is important to bear in mind that the present theory is not designed to deal with very large levels of unsteadiness. For this reason, it does not in general provide the exact form of the kernel at very short times, {\color{black}{i.e. $t\lesssim\epsilon^2$}}. Another example of this limitation is observed in the case of the force acting on a spherical bubble, for which the total contribution proportional to the instantaneous relative acceleration is found to be the sum of the added-mass force and of a (dominant) term left by the early decay of the exact history kernel. In the present theory, history effects appear at first and second orders through the far-field velocity corrections $\vec{\mathcal{U}}_1(t)$ and $\vec{\mathcal{U}}_2(t)$, respectively. While these outer contributions are crucial in the inertial corrections to the force, they do not affect the torque in the case of a spherical (or spheroidal) particle, for symmetry reasons. \\
\indent In (\ref{Torque}), two quadratic contributions to the second-order inertial torque are identified. We showed that the one proportional to $\ma{S}\cdot\boldsymbol\Omega$ is identical to that computed by \cite{candelier2016angular} for a neutrally buoyant particle. It is nonzero only in three-dimensional linear flows whose structure generates a uniform vortex stretching, which makes it relevant to the motion of small particles in turbulence. The contribution proportional to $\ma{S}\cdot\boldsymbol\omega_s$ was apparently not identified so far. It reveals that a spinning particle immersed in a linear flow with a nonzero strain component aligned with the spin axis experiences an inertial torque. This correction makes the total torque $\boldsymbol\tau'=\boldsymbol\tau'^{(0)}+\epsilon^2\boldsymbol\tau'^{(2)}$ weaker (resp. stronger) than the primary viscous torque if the particle spins about the elongational (resp. compressional) axis of the straining flow. It induces a torque component orthogonal to the primary spinning direction otherwise. For instance, the total torque on a particle spinning with the angular velocity $\omega_p\vec{e}_1$ in the linear shear flow $\vec{U}^\infty=x_3\vec{e}_1$ is $\boldsymbol\tau'=-8\pi\omega_p(\vec{e}_1-\frac{\epsilon^2}{16}\vec{e}_3)$.\\
\indent The second-order force (\ref{Force}) involves three quadratic contributions. The one corresponding to the spin-induced lift force originally computed by \cite{rubinow1961transverse} is well known. Our results show that this force subsists in a linear flow, provided the slip rotation rate $\boldsymbol{\omega}_s=\boldsymbol{\omega}_p-\boldsymbol\Omega$ is substituted for the particle spin rate $\boldsymbol{\omega}_p$.  The other two contributions involve the kinematic structure of the background flow, through $\ma{S}$ and $\boldsymbol\Omega$, and the particle slip velocity, $\vec{u}_s$.  The contribution proportional to $\boldsymbol\Omega\times\vec{u}_s$ is a pure lift force while that proportional to $\ma{S}\cdot\vec{u}_s$ may comprise drag and lift components. We compared the pre-factors of the corresponding terms with those found in inviscid flow in the weak shear limit ($as/u_c\ll1$). We also compared the pre-factors of the $\ma{S}\cdot\vec{u}_s$ contribution determined in an axisymmetric uniaxial flow for a rigid particle and a spherical bubble, respectively. These comparisons shed light on the central role played by the dynamic boundary condition at the particle surface. To a large extent, the relative magnitude of these contributions is determined by the changes imposed to the vorticity distribution at the particle surface by the strain or/and rotation component of the base flow. These changes modify the spatial distribution of the fluid momentum in the region close to the particle, i.e. at distances less than $\ell_s$, which in turn results in a net force on it. The velocity gradients $\ma{S}$ and $\boldsymbol\Omega$ enter the kernel $\ma{K}$ in a subtle manner (this is what makes it flow-dependent), hence also the far-field velocity correction $\vec{\mathcal{U}}_2$. For a given flow, the second-order force on a rigid particle tends to $\frac{37\pi}{12} \ma{S} \cdot \vec{u}_s + \frac{4\pi}{3} \boldsymbol{\Omega}\times \vec{u}_s+ 6\pi \vec{\mathcal{U}}_2$ in the long-term limit. We found that, for a given slip velocity, the inner and outer contributions have opposite signs, and the sign of their sum depends on the considered flow, as the examination of several canonical configurations in \S\,\ref{disc} revealed. Moreover, the total lift force is obtained by summing the first- and second-order lift components. The two may have opposite signs, especially in vortical flows. Although the first-order term is necessarily dominant from an asymptotic point of view, we observed that in a solid-body rotation flow the pre-factor of the second-order component is significantly larger than that of the first-order one, making the total lift force change sign beyond a modest value of the shear Reynolds number.\\
\indent Having isolated the $\frac{37\pi}{12} \ma{S} \cdot \vec{u}_s$ contribution to the second-order force in straining flows led us to re-examine the old yet still open question whether one should use the Lagrangian derivative, $\frac{\mbox{D} \vec{U}^\infty}{\mbox{D} t}\Big|_{\vec{x}_p}$, or the time derivative following the particle, $\frac{\mbox{d} \vec{U}^\infty}{\mbox{d} t}\Big|_{\vec{x}_p}$, in the expression for the added-mass force in nonuniform flows. Indeed, selecting the former, which is known to be the proper choice in the inviscid limit, introduces a contribution $-\frac{2\pi}{3}\ma{S} \cdot \vec{u}_s$ in the second-order force in straining flows, while no such contribution appears if the latter is chosen. However, the above question turned out to be undecidable at the present order of approximation, owing to the dominant influence of the dynamic boundary condition on the strength of the $\ma{S} \cdot \vec{u}_s$ term, as underlined by the opposite signs of the above two pre-factors. Indeed, the flow configurations examined in \S\,\ref{disc} did not bring any convincing clue that might allow us to conclude that the $\frac{37\pi}{12}$ pre-factor should rather be interpreted as $-\frac{2\pi}{3}+\frac{15\pi}{4}$, with the first and second contributions corresponding respectively to the added-mass effect induced by the no-penetration condition and the `vortical' effect resulting from the dynamic boundary condition. Higher-order expansions, or more likely DNS results covering a wide range of Reynolds number, such as those of \cite{Magnaudet1995}, may help to disentangle the two types of contributions. At the moment, the latter reference provides the clearest heuristic evidence that the inviscid form of the added-mass term remains valid at finite Reynolds number.\vspace{2mm}\\
\indent In view of applications, it appears useful to summarize the complete expressions for the force and torque resulting from the present theory in dimensional variables. {\color{black}{We remind the reader that this theory assumes that the background flow has the form $\vec{U}_\infty(\vec{x},t)=\vec{U}_0(t)+\ma{S}\cdot\vec{x}+\boldsymbol\Omega\times\vec{x}$ (hence $\boldsymbol{\nabla} \times \vec{U}^\infty=2\boldsymbol\Omega$), with $\ma{S}$ and $\boldsymbol\Omega$ uniform and time-independent. Actually, the latter restriction applies to the results for the force, while those for the torque remain valid even through $\ma{S}$ and $\boldsymbol\Omega$ depend upon time. This restriction also implies that, in three-dimensional flows, the results for the force are only valid in pure straining flows ($\boldsymbol\Omega=\vec{0}$). With these restrictions in mind,}} using the definitions introduced in \S\S\,1 and 2, the total force and torque, obtained by summing the creeping-flow result with the first- and second-order inertial corrections and the force/torque resulting from the undisturbed flow, read
\begin{eqnarray}
\nonumber
\label{statd2}
\vec{f}(t)&=&\vec{f}_b(t)-6\pi\mu a\left\{\vec{u}_s+\Big(\frac{a^2s}{\nu}\Big)^{1/2}  \displaystyle{\int_0^t} 6\pi \ma{K}\big(s(t-\tau)\big)\cdot \frac{\text{d}\vec{u}_s}{\text{d}\tau}d\tau\right\}\\
\nonumber
&+&m_f\Biggl\{4 \frac{\mbox{d}\vec{u}_s}{\mbox{d}t} +\frac{37}{16} \ma{S} \cdot \vec{u}_s 
-  \frac{1}{2} \vec{u}_s \times \left( \boldsymbol{\nabla} \times \vec{U}^\infty\right) 
- \frac{3}{4}
\vec{u}_s\times\boldsymbol{\omega}_s\Biggr.  \\
  \Biggl. &-& \frac{9}{2} \displaystyle{\int_0^t} 6\pi \ma{K}\big(s(t-\tau)\big)\cdot  \frac{\text{d}}{\text{d}\tau}\Big( \int_0^\tau 6\pi \ma{K}\big(s(t-\tau)\big)\cdot \frac{\text{d}\vec{u}_s}{\text{d}\tau'}d\tau'\Big)\text{d}\tau\Biggr\}\,, \\
 \boldsymbol{\tau}(t) &=&-8\pi\mu a^3\boldsymbol{\omega}_s -m_fa^2\left(2\frac{\mbox{d}\boldsymbol{\omega}_s }{\mbox{d}t} +
  \frac{6}{5}\ma{S}\cdot \left( \boldsymbol{\nabla} \times \vec{U}^\infty\right)  -\frac{3}{4}\ma{S}\cdot \boldsymbol{\omega}_s\right)\,,
  \label{Torqued}
\end{eqnarray}
with 
\begin{eqnarray}
\nonumber
\vec{f}_b(t)&=&(m_p-m_f)\vec{g}+m_f\frac{\mbox{D} \vec{U}^\infty}{\mbox{D} t} \Big|_{\vec{x}_p(t)}\\
&=&(m_p-m_f)\vec{g}+m_f\left\{\dot{\vec{U}}_0+\ma{S}\cdot\vec{U}^\infty\big|_{\vec{x}_p(t)}+\frac{1}{2}\left( \boldsymbol{\nabla} \times \vec{U}^\infty\right) \times\vec{U}^\infty\big|_{\vec{x}_p(t)}\right\}  \,.
\end{eqnarray}
Since $\ma{K}(st)\rightarrow\frac{1}{6\pi}\frac{\ma{I}}{\sqrt{\pi st}}$ for $st\rightarrow0$,
 the dimensional force in the short-term limit is
\begin{eqnarray}
\label{statd1}
\vec{f}({t\ll s^{-1}})&=&\vec{f}_b(t)-6\pi\mu a\left\{\vec{u}_s+\int_0^t \frac{1}{\sqrt{\pi\nu(t-\tau)/a^2}} \frac{\text{d}\vec{u}_s}{\text{d}\tau}d\tau\right\}\\
\nonumber
&&+m_f\Biggl\{-\frac{1}{2} \frac{\mbox{d}\vec{u}_s}{\mbox{d}t} +\frac{37}{16} \ma{S} \cdot \vec{u}_s 
-  \frac{1}{2} \vec{u}_s \times \left( \boldsymbol{\nabla} \times \vec{U}^\infty\right) 
- \frac{3}{4}
\vec{u}_s\times\boldsymbol{\omega}_s\Biggr\}  \,.
\end{eqnarray}
This is the original BBO equation (\ref{eq:bbo}) enriched with the last three terms which account for the interaction of the slip velocity with the background velocity gradients and the particle rotation. {\color{black}{Note that (\ref{statd1}) remains valid for times of $\mathcal{O}(a^2/\nu)$ or even shorter, since the short-time limit of the history kernel and that of the added-mass force coincide with those found in the BBO approximation. This is a bonus specific to the case of the force on a rigid spherical particle. 
The same does not hold for a bubble, nor for the torque on a spherical particle, because in these cases the BBO kernel involves a term proportional to $\exp[{9\nu(t-\tau)/a^2}]\text{erfc}[{9\nu(t-\tau)/a^2}]^{1/2}$ which is not captured by the present theory. 
A similar term is known to exist for drops of arbitrary viscosity \citep{Gorodtsov1975,Yang1991} and nonspherical particles \citep{Lawrence1986}. Obtaining expressions for the loads that incorporate finite-Reynolds-number effects and are uniformly valid in the limit $t\rightarrow0$ even when the exact kernel does not reduce to the Basset-Boussinesq form $[\nu(t-\tau)/a^2]^{-1/2}$ requires a theory fundamentally different from the one described here, the starting point of which is the unsteady Stokes equation. Since, instead of the simple form (\ref{O0}), the solution to this equation is nonlocal in time, a good part of the methodology outlined in \S\,\ref{sec:sol} no longer applies and it is doubtful that results may be obtained in closed form in the time domain. Nevertheless, the problem is worthy of consideration since in emerging fields, such as acoustofluidics, the particle motion is driven by a high-frequency acoustic field, which corresponds to high levels of unsteadiness with $\text{Re}_s\text{Sl}=\mathcal{O}(1)$ in (\ref{eq:outer_tp}), while small-but-finite advective effects have a significant influence \citep{Agarwal2021}. }} \vspace{2mm}\\
\indent For $st\rightarrow\infty$, $\ma{K}(st)\rightarrow \overline{\ma{K}}$, so that 
in the long-term limit one has
\begin{eqnarray}
\label{shortd1}
\vec{f}({t\gg s^{-1}})&=&\vec{f}_b-6\pi\mu a\left\{\vec{u}_s+\Big(\frac{a^2s}{\nu}\Big)^{1/2} (6\pi \overline{\ma{K}}) \cdot \vec{u}_s\right\}\\
\nonumber
&+&m_f\Biggl\{\frac{37}{16} \ma{S} \cdot \vec{u}_s 
-  \frac{1}{2} \vec{u}_s \times \left( \boldsymbol{\nabla} \times \vec{U}^\infty\right) 
- \frac{3}{4}
\vec{u}_s\times\boldsymbol{\omega}_s    - \frac{9}{2}s(6\pi \overline{\ma{K}}) \cdot (6\pi \overline{\ma{K}}) \cdot \vec{u}_s\Biggr\} \,.
\end{eqnarray}
In (\ref{statd2})-(\ref{shortd1}) the velocity-gradient scale $s$ is defined as $s=(\frac{1}{\mathcal{D}!}\ma{A}\colon\ma{A})^{1/2}$, with $\mathcal{D}=1,\,2$ or $3$ depending on whether the base flow is one-, two- or three-dimensional.\vspace{2mm}\\
\indent Besides the fact that the background flow is assumed stationary and all Reynolds numbers involved have to be small, the present theory suffers from two main limitations. First, the condition $\text{Re}_p\ll\text{Re}_s^{1/2}$ requires the slip velocity to be very small, making the theory essentially suitable for weakly positively or negatively neutrally buoyant particles. In particular, our second-order results do not comprise the classical Oseen correction to the drag, which results from advective terms that are negligible in the outer region under the above condition. Beyond the drag increase they induce \citep{Kaplun1957,Proudman1957}, these terms are known to decrease the strength of the Saffman lift force \citep{asmolov1990,mclaughlin1991inertial}. For instance, this force is reduced by $25\%$ when $\text{Re}_p\approx\text{Re}_s^{1/2}$, and the larger the ratio $\text{Re}_p/\text{Re}_s^{1/2}$ the smaller the lift force. Consequently, finite slip effects affect the kernel components involved in the first- and second-order inertial corrections, and extending our theory to incorporate these effects is crucial to make it applicable to situations involving large fluid-to-particle density ratios. \\
\indent Second, the kernel is only known for the simplest linear flows. Since the problem is non-linear, one cannot just determine the general kernel by linearly superposing the \lq elementary\rq{} kernels for the canonical flows that were considered in \S\,4. 
To make the present theory useful in applications involving arbitrary linear flows, it is necessary to render the dependence of the kernel with respect to the components of the velocity-gradient tensor $\ma{A}$ explicit. Closed-form expressions of $\ma{K}$ in a general linear flow may presumably be obtained only in the short- and long-term limits, while semi-empirical fits will certainly be required to obtain approximate expressions valid for arbitrary times. This, we think, is the price to pay for making the results (\ref{statd2})-(\ref{shortd1}) applicable in practice to a wide range of configurations of interest in sheared suspensions and possibly in turbulent flows.
 \vspace{-5mm}
 \section*{ Acknowledgements}
 \noindent BM was supported by Vetenskapsr\r{a}det (grant no. 2021-4452) and by the Knut-and-Alice Wallenberg Foundation (grant no. 2019.0079). 
 \section*{Declaration of interests.}  
 \noindent The authors report no conflict of interest.
\appendix

 \section{Second-order force on a spherical bubble}
 \label{bubble}

 \label{AppendixA}
We complemented the results for a rigid sphere with those for a spherical bubble. Our initial hope was that this might allow us to disentangle inertial effects resulting from the particle shape (i.e., added-mass effects) from those resulting from the vorticity generated at the particle surface. Indeed, comparing a spherical bubble to a rigid sphere
leaves the no-penetration condition unchanged, whereas the no-slip condition is changed into a shear-free condition for the bubble, assuming that the viscosity of the gas that fills it is negligibly small and the gas-liquid interface is free of any contamination. 
For technical reasons related to the determination of the second-order inner solution, we did not succeed to solve this problem for a general linear flow. Nevertheless, we managed to solve it when the bubble moves, with a possibly time-dependent slip velocity, in the axisymmetric straining flow considered in appendix \ref{uniax}.\\ 
\indent The steps involved in the calculation of the first- and second-order force disturbance on a spherical bubble are similar to those described in \S\,\ref{sec:sol}. Only the boundary condition at the particle surface differs. In the $\mathcal{O}(\epsilon)$-problem, the first boundary condition in (\ref{BC:v_in}) is replaced by 
\begin{equation}
{\vec{v}}_{\mbox{\scriptsize in}}^{(1)} \cdot \vec{n} =    -\vec{\mathcal{U}}^{(1)}(t) \cdot \vec{n}\,,\quad 
\vec{n}\times(\boldsymbol\sigma^{(1)}\cdot \vec{n})= \vec{0} 
\: \quad \mbox{at}\quad r=1\,,
\label{vn}
\end{equation}
 while at order $\epsilon^2$, instead of the first boundary condition in (\ref{BC:w_h}) one has 
 \begin{equation}
 \vec{w}_{h} \cdot \vec{n} = - \left(\vec{w}_p + \vec{\mathcal{U}}_2(t) \right) \cdot \vec{n}\:, \quad\quad \vec{n}\times(\boldsymbol\sigma_{h}\cdot \vec{n})= - \vec{n}\times(\boldsymbol\sigma_{p}\cdot \vec{n})\,\quad\mbox{at}\quad   r=1 \:.
\label{taun}
 \end{equation}
 In (\ref{vn}), $\boldsymbol\sigma^{(1)}\cdot\vec{n}$ denotes the first-order viscous traction resulting from the velocity field ${\vec{v}}_{\mbox{\scriptsize in}}^{(1)}$, while in (\ref{taun}), $\sigma_{h}\cdot \vec{n}$ and $\sigma_{p}\cdot \vec{n}$ denote the second-order viscous tractions resulting from the velocity components $\vec{w}_{h}$ and $\vec{w}_{p}$ defined in (\ref{ph}), respectively.
 
 Carrying out the steps described in \S\,\ref{sec:sol}, the total disturbance force acting on the bubble is found to be
 \begin{equation}
\begin{split}
\vec{f}'&= \vec{f}'^{(0)} +\epsilon\vec{f}'^{(1)}+\epsilon^2\vec{f}'^{(2)}=   -4\pi \vec{u}_s - 4\pi \epsilon \int_0^t \ma{K}(t-\tau)  \frac{\mbox{d} \vec{f}^{(0)}}{\mbox{d}\tau} \mbox{d} \tau \\
 &+ \epsilon^2 \left(2\pi \frac{\mbox{d} \vec{u}_s}{\mbox{d}t}+\frac{4\pi}{3} \ma{S} \cdot \vec{u}_s  - 4\pi \int_0^t \ma{K}(t-\tau) \frac{\mbox{d}}{\mbox{d}\tau}\left[ \int_0^\tau 4\pi \ma{K}(\tau-\tau') \frac{\mbox{d} \vec{f}^{(0)}}{\mbox{d}\tau'} \mbox{d}\tau'\right] \mbox{d}\tau \right)\:.
\end{split}
\label{Force_bubble}
\end{equation}
\begin{table}
\label{table2}
\caption{Summary of all contributions
to the second-order force on a bubble moving along the axis of an axisymmetric elongational flow, expressed with respect to $\vec{f}^{(0)} = 4\pi\vec{u}_s$. Terms arising from the regular and singular parts of the solution are given in the upper and lower tables, respectively. In the last row of the former, note that $\frac{3029}{6200}  -\frac{281}{9300}-\frac{1}{8}=\frac{1}{3}$, which yields the $\frac{4\pi}{3}\ma{S}\cdot\vec{u}_s$ contribution in (\ref{Force_bubble}).
We also distinguish between the contributions to the force induced by the viscous part of the stress tensor (e.g. $\vec{f}'_\nu$)  and 
those induced by the pressure (e.g. $\vec{f}'_p$).  
The total second-order hydrodynamic force is the sum of these two contributions.
}


\mbox{}\\[1cm]
\centering {\bf Regular terms}
$$
\begin{array}{c | c | c | c | c | c }
\hline 
 & 
  & & &\\
 & \partial_t \vec{w}_{\mbox{\scriptsize in}}^{(0)} & \ma{S}\cdot\vec{w}_{\mbox{\scriptsize in}}^{(0)} + (\ma{S}\cdot \vec{r}) \cdot \boldsymbol{\nabla} \vec{w}_{\mbox{\scriptsize in}}^{(0)} & 
 -\vec{u}_s \cdot \boldsymbol{\nabla} \vec{w}_{\mbox{\scriptsize in}}^{(0)}  &  \vec{w}_{\mbox{\scriptsize in}}^{(0)} \cdot \boldsymbol{\nabla}\vec{w}_{\mbox{\scriptsize in}}^{(0)} \\
 &  
  & & &\\
\hline 
\hline
 & 
  & & &\\
\vec{f}'^{(2)}_\nu 
& \frac{4}{9} \frac{\text{d} \vec{f}^{(0)}}{\text{d}t} & \frac{803}{3100} \ma{S} \cdot \vec{f}^{(0)} & \frac{113}{4650} \ma{S} \cdot \vec{f}^{(0)} & -\frac{1}{20} \ma{S} \cdot \vec{f}^{(0)} \\
 &  
 & & &\\
\hline 
 & 
  & & &\\
 \vec{f}'^{(2)}_p  
 & \frac{1}{18} \frac{\text{d} \vec{f}^{(0)}}{\text{d}t}  & \frac{1423}{6200} \ma{S} \cdot \vec{f}^{(0)}& -\frac{169}{3100} \ma{S} \cdot \vec{f}^{(0)}& -\frac{3}{40} \ma{S} \cdot \vec{f}^{(0)}\\
&  
& & &\\
\hline 
& 
& & &\\
 \vec{f}'^{(2)}
 & \frac{1}{2}\frac{\text{d} \vec{f}^{(0)}}{\text{d}t}  & \frac{3029}{6200} \ma{S} \cdot \vec{f}^{(0)}&  -\frac{281}{9300} \ma{S} \cdot \vec{f}^{(0)}  & -\frac{1}{8}\ma{S} \cdot \vec{f}^{(0)}\\
&   
 & & &\\
\hline 
\end{array}
$$
\mbox{}\\[1cm]
\centering {\bf Singular terms,} with 
$\quad \vec{\mathcal{U}}_2(t) = -  \int_0^{t} \ma{K}(t-\tau) \cdot  \frac{\mbox{d}}{\mbox{d}\tau}\left( \int_0^{\tau} 4\pi \:\ma{K}(\tau-\tau') \cdot \frac{\mbox{d}\vec{f}^{(0)}(\tau')}{\mbox{d}\tau'} \mbox{d}\tau' \right) \mbox{d} \tau $
$$
\begin{array}{c | c | c | c | }
\hline 
 & & & \\
& 
 & \mbox{Short-time behaviour: } \displaystyle{\ma{K}(t) \sim \frac{1}{6\pi}\frac{1}{\sqrt{\pi t}}} &\mbox{Long-time behaviour: } \displaystyle{\ma{K}(t) \to \overline{\ma{K}}}   \\  
  & & &\\
\hline 
\hline 
  & & &\\
\vec{f}'^{(2)}_\nu & \frac{8\pi}{3} \vec{\mathcal{U}}_2(t) 
& -\frac{8}{27} \frac{\text{d} \vec{f}^{(0)}}{\text{d}t}  & - \frac{8\pi}{3}\:\overline{\ma{K}}\cdot (4\pi \overline{\ma{K}})\cdot \vec{f}^{(0)} \\
 & & &\\
\hline 
  & & &\\
 \vec{f}'^{(2)}_p  
 & \frac{4\pi}{3}  \vec{\mathcal{U}}_2(t)  
 & - \frac{4}{27} \frac{\text{d} \vec{f}^{(0)}}{\text{d}t} & - \frac{4\pi}{3} \:\overline{\ma{K}}\cdot (4\pi \overline{\ma{K}})\cdot \vec{f}^{(0)} \\ 
& & &\\
\hline 
& & &\\
 \vec{f}'^{(2)}
& 4\pi \vec{\mathcal{U}}_2(t) 
& -\frac{4}{9}\frac{\text{d} \vec{f}^{(0)}}{\text{d}t} & - 4\pi\:\overline{\ma{K}}\cdot (4\pi \overline{\ma{K}})\cdot \vec{f}^{(0)} \\
 & & &\\
\hline 
\hline 
\end{array}
$$
\label{Tab2}
\end{table}

As for a rigid sphere, the different  terms in (\ref{Force_bubble}) may be split into pressure and viscous contributions. The result is detailed in table \ref{Tab2}. Note that the $\frac{4\pi}{3} \ma{S} \cdot \vec{u}_s$ contribution was only determined for a bubble located on the symmetry axis of the axisymmetric straining flow defined by (\ref{s3d}). However, whatever the bubble position and the velocity-gradient tensor $\ma{A}$ characterizing the linear straining flow, the stationary second-order force resulting from the inner solution must have the vector form $\ma{S} \cdot \vec{u}_s$, similar to (\ref{Force}) for a rigid sphere. This is why the particular result derived for a bubble on the symmetry axis of a uniaxial flow remains valid in an arbitrary linear flow. \\
\indent Apart from the fact that the pre-factors $6\pi$ are replaced by $4\pi$ for a bubble, (\ref{Force_bubble}) differs from (\ref{Force}) through the pre-factors of the second-order contributions  $\frac{\mbox{d} \vec{u}_s}{\mbox{d}t}$ ($2\pi$ instead of $\frac{16\pi}{3}$) and $\ma{S} \cdot \vec{u}_s$ ($\frac{4\pi}{3}$ instead of $\frac{37\pi}{12}$). The significance of the latter difference is discussed in \S\,\ref{disc}. In what follows we examine the origin of the former. Indeed, once the inner solution and the short-time contribution of the kernel are combined, the $\frac{\mbox{d} \vec{u}_s}{\mbox{d}t}$ term should provide the same added-mass force $-  \frac{2\pi}{3} \epsilon^2\frac{\mbox{d} \vec{u}_s}{\mbox{d}t}$ as that found in (\ref{Force2}) for a rigid sphere, the added-mass force being independent of the dynamic boundary condition. \\
 \indent Since the kernel depends on the boundary condition at the particle surface only through the pre-factors $6\pi$ or $4\pi$, one still has $6 \pi\ma{K}(t) = \frac{\ma{I}}{\sqrt{\pi t}}$ at short times. Inserting this asymptotic expression in (\ref{Force_bubble}) yields
\begin{equation}
\vec{f}'(\epsilon^2\ll t\ll1)  = -4\pi \vec{u}_s - \frac{8\pi}{3}   \epsilon \:\int_0^t  \frac{1}{\sqrt{\pi (t-\tau)}} \frac{\mbox{d} \vec{u}_s}{\mbox{d}\tau} \mbox{d} \tau + \epsilon^2\big( \frac{2 \pi}{9} \frac{\mbox{d} \vec{u}_s}{\mbox{d}t}+\frac{4\pi}{3} \ma{S} \cdot \vec{u}_s\big) \:.
\label{Force_bubble_2}
\end{equation}
\vspace{2mm}\\
Surprisingly, the pre-factor of the $\frac{\mbox{d} \vec{u}_s}{\mbox{d}t}$ term is $\frac{2 \pi}{9}$ instead of $-\frac{2\pi}{3}$ as expected. To understand this difference, one has to compare (\ref{Force_bubble_2}) (considering from now on $\ma{S}=\vec{0}$) with the exact solution obtained by \cite{Gorodtsov1975} and \cite{Yang1991} for a bubble translating unsteadily in a fluid at rest in the limit of negligibly small advective effects, namely
\begin{equation}
\vec{f}_{\rm G} = -4\pi \Biggl\{\vec{u}_s +2\int_0^t \exp \left( \frac{9 \:(t-\tau)}{\epsilon^2}\right) \mbox{erfc}\left( \sqrt{\frac{9 (t-\tau)}{\epsilon^2}}\right) \frac{\mbox{d} \vec{u}_s}{\mbox{d}\tau} \mbox{d} \tau \Biggr\}-  \frac{2\pi}{3} \epsilon^2\frac{\mbox{d} \vec{u}_s}{\mbox{d}t}\:.
\label{GY}
\end{equation}
Similar to the BBO equation (\ref{BBO}), the three terms on the right-hand side respectively represent the quasi-steady drag, the visco-inertial history effect and the added-mass force, the latter resulting exclusively from the pressure contribution to the total force. 
At first glance, (\ref{GY}) confirms that the second-order prediction (\ref{Force_bubble_2}) differs from the exact result. However, the present theory is designed to work up to large but `not too large'  levels of unsteadiness. In particular, it is only supposed to work for times much larger than $\epsilon^2$. In this limit, the kernel involved in (\ref{GY}) is such that, at leading order,
\begin{equation}
\exp \left( \frac{9 \:t}{\epsilon^2}\right) \mbox{erfc}\left( \sqrt{\frac{9 t}{\epsilon^2}}\right)  \sim  \frac{1}{3}  \frac{\epsilon }{\sqrt{\pi t}} \,.
\label{exp1}
\end{equation}

To find out the next term in the expansion of this kernel for large $t/\epsilon^2$, one may start from the mathematical result $\int_0^\infty\left\{\frac{1}{\sqrt{\pi x}}-\exp \left(  x\right) \mbox{erfc}\left( \sqrt{ x}\right) \right\}dx=1$. Moreover, in the sense of generalized functions, one knows that $1=\int_{-\infty}^{+\infty}\delta(t)dt$, with $\delta(t)$ the Dirac delta function. Then, setting $x=t/\zeta$ and defining the function $F_\zeta (t)$ such that $F_\zeta (t)\equiv0$ for $t<0$ and $F_\zeta (t)=\frac{1}{\zeta}\left\{\frac{1}{\sqrt{\pi t/\zeta}}-\exp(  t/\zeta) \mbox{erfc}( \sqrt{ t/\zeta})\right\}$ for $t>0$, one finds (\cite{Boccara}, pp. 53-54)
\begin{equation}
\mbox{lim}_{\zeta\rightarrow0}F_\zeta (t)=\delta(t)\,.
\label{asymdelta}
\end{equation}
Finally, setting $\zeta=\epsilon^2/9$, one obtains

\begin{equation}
\lim_{t/\epsilon^2 \gg 1} \exp \left( \frac{9 \:t}{\epsilon^2}\right) \mbox{erfc}\left( \sqrt{\frac{9 t}{\epsilon^2}}\right) =\frac{1}{3} \epsilon \frac{1}{\sqrt{\pi t}} -\frac{1}{9}\epsilon^2 \delta(t)+\mathcal{O}(\epsilon^3)\:. 
\label{GYlim}
\end{equation}
Inserting this result into (\ref{GY}), one recovers (\ref{Force_bubble_2}). 
Therefore, in contrast to the case of a rigid sphere, the history force acting on a bubble contains a term of $\mathcal{O}(\epsilon^2)$. For this reason, after a very short transient, the $\mathcal{O}(\epsilon^2)$-correction to the force acting on the bubble results from a combination of added mass and visco-inertial history effects.  The latter is even dominant since the resulting pre-factor, $\frac{2\pi}{9}$, has an opposite sign to that of the added-mass force, $-\frac{2\pi}{3}$. Similar to the problem encountered with the terms $\ma{S}\cdot\vec{u}_s$ and $\boldsymbol\Omega\times\vec{u}_s$ in \S\,\ref{disc}, one faces a situation is which, under a certain asymptotic condition, a term with a given mathematical structure encapsulates two effects produced by two distinct physical mechanisms. 
 A similar situation was identified by \cite{Magnaudet2003} for a rigid particle moving unsteadily close to a wall: also in this case, the contribution proportional to the instantaneous relative acceleration between the particle and fluid results from a combination of added-mass and history effects (see equation (14) of this reference and the discussion that follows this equation).

\begin{figure}
\vspace{4mm}
\begin{center}
\includegraphics[width=0.7\linewidth]{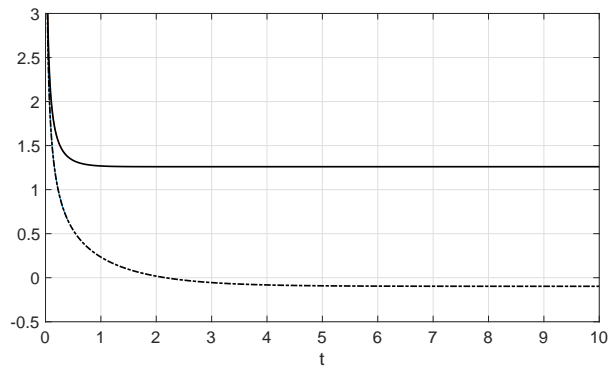}
\end{center}
\caption{Evolution of the kernel $\ma{K}(t)$ in a uniaxial straining flow. Solid line: axial component $\ma{K}_3^3$; dotted line: radial components $\ma{K}_1^1=\ma{K}_2^2$. }
\label{kernelaxi}
\end{figure}\noindent 
  \section{Kernel in an axisymmetric straining flow}
  \label{uniax}
  In CMM, only planar linear flows were considered. However another configuration of fundamental interest is the so-called uniaxial straining flow corresponding to the velocity-gradient tensor
 \begin{equation}
\ma{A} = \left(\begin{array}{ccc} -1 & 0 & 0\\
0 & -1 & 0 \\
0 & 0 & 2\\
\end{array}\right) \:.
\label{s3d}
\end{equation}

The forces experienced by a spherical rigid sphere or a bubble held fixed in this flow were considered numerically by \cite{Magnaudet1995}. We computed the kernel corresponding to this flow using the techniques described by CMM. Figure \ref{kernelaxi} shows how the nonzero components of $\ma{K}$ reach their stationary value. 
At short times, these components behave as
\begin{equation}
6\pi \ma{K}(t) \sim  \frac{1}{\sqrt{\pi t}}\left(\begin{array}{ccc}
 1- \frac{7}{10 } t +\frac{2}{21 }t^{2}...    
 & 0 & 0 \\
0 &    1- \frac{7}{10 } t +\frac{2}{21 }t^{2}...     & 0\\
0 & 0 &  1 + \frac{7}{5 } t  -  \frac{19}{210 }t^{2}...   \\
\end{array}\right) \,.
\label{kt3d}
\end{equation}
That is, deviations from the Basset-Boussinesq limit grow initially as $t^{1/2}$. Up to terms of $\mathcal{O}(t^{1/2})$, the short-time evolution of the kernel may be written in the simple form $6\pi \ma{K}(t) \sim  \frac{1}{\sqrt{\pi t}}(\ma{I}+\frac{7}{10}\ma{S}t+...)$. This form was shown to be universal, i.e. independent of the specific linear flow considered, by \cite{Bedeaux1987} and \cite{miyazaki1995drag}. According to figure \ref{kernelaxi}, the axial and radial components reach their long-term asymptote at $t\approx1$ and $t\approx4$, respectively. Interestingly, the radial component changes sign beyond $t\approx2.2$. This means that the long-time $\mathcal{O}(\epsilon)$-correction to the drag is negative, so that the drag of a particle translating in the radial direction (along which fluid elements are compressed) is decreased compared to its creeping-flow value. In the stationary regime, one has
\begin{figure}
\vspace{-30mm}
\begin{center}
\includegraphics[width=0.7\linewidth]{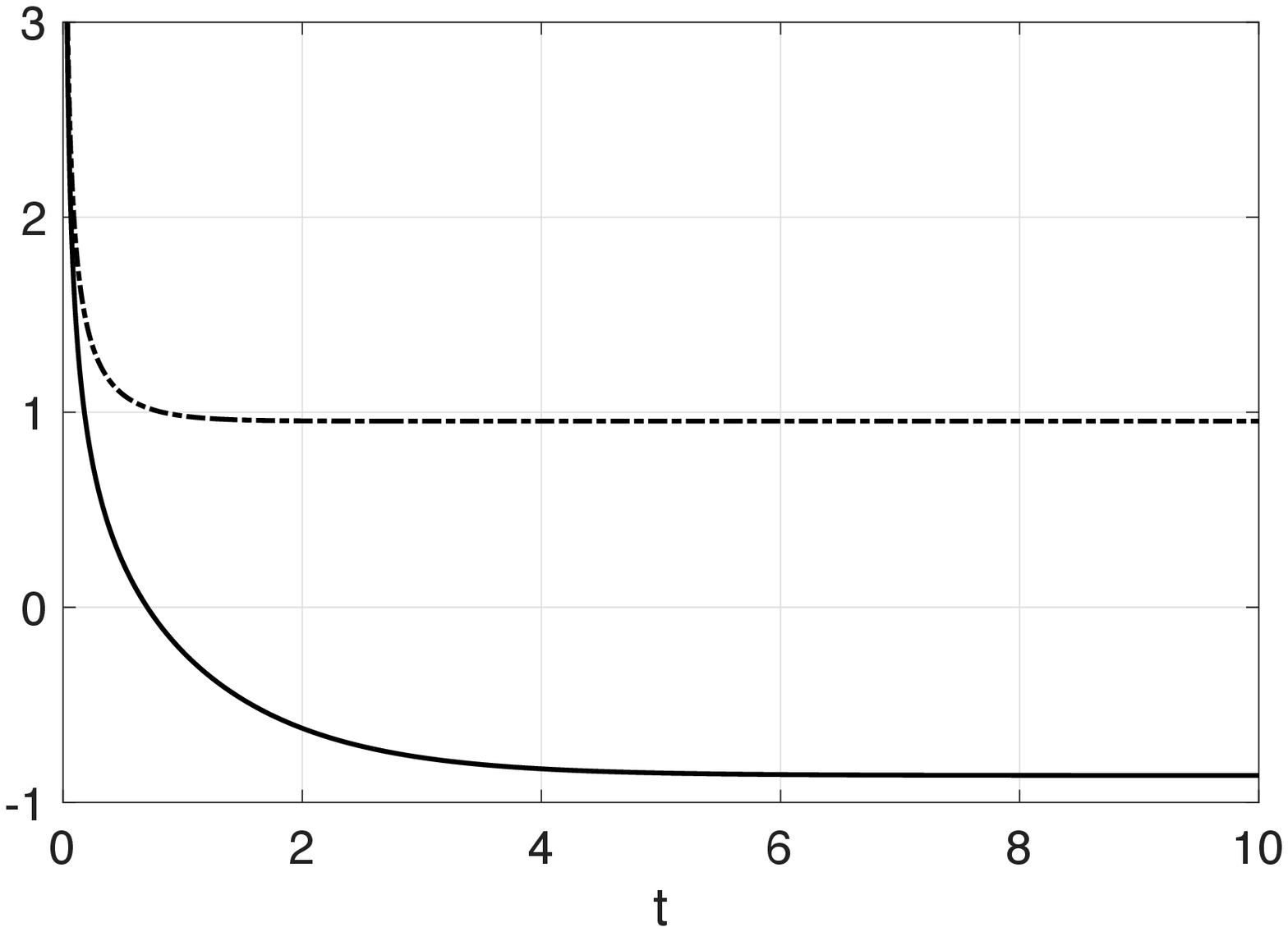}
\end{center}
\vspace{-35mm}
\caption{Evolution of the kernel $\ma{K}(t)$ in a biaxial straining flow. Solid line: axial component $\ma{K}_3^3$; dotted line: radial components $\ma{K}_1^1=\ma{K}_2^2$. }
\label{kernelaxi2}
\end{figure}
\begin{equation}
6\pi\overline{\ma{K}} \approx\left(\begin{array}{ccc} 
 -0.097 & 0 & 0 \\
0 & -0.097 & 0\\
0 & 0 & 1.260 \\
\end{array}\right) \:.
\label{kbar3d}
\end{equation}
Since the $\mathcal{O}(\epsilon^2)$-correction to the drag is proportional to $\overline{\ma{K}}\cdot\overline{\ma{K}}$, this correction is found to be significant along the flow axis but negligibly small in the radial direction.\vspace{2mm}\\
{\color{black}{\indent We also computed the kernel associated with the so-called biaxial straining flow associated with the velocity-gradient tensor
 \begin{equation}
\ma{A} = \left(\begin{array}{ccc} 1 & 0 & 0\\
0 & 1 & 0 \\
0 & 0 & -2\\
\end{array}\right) \:.
\label{s3d2}
\end{equation}
At short times, the components of this kernel behave as
\begin{equation}
6\pi \ma{K}(t) \sim  \frac{1}{\sqrt{\pi t}}\left(\begin{array}{ccc}
 1+ \frac{7}{10 } t +\frac{2}{21 }t^{2}...    
 & 0 & 0 \\
0 &    1+ \frac{7}{10 } t +\frac{2}{21 }t^{2}...     & 0\\
0 & 0 &  1 - \frac{7}{5 } t  -  \frac{19}{210 }t^{2}...   \\
\end{array}\right) \,.
\label{kt3d2}
\end{equation}
Comparing (\ref{kt3d}) and (\ref{kt3d2}) indicates that even-order terms are identical while odd-order terms have the same magnitude but opposite signs, as expected from the universal short-time form of $\ma{K}(t)$ mentioned above. \\
\indent The corresponding steady-state kernel is found to be
\begin{equation}
6\pi\overline{\ma{K}} \approx\left(\begin{array}{ccc} 
 0.9544 & 0 & 0 \\
0 & 0.9544 & 0\\
0 & 0 & -0.8622 \\
\end{array}\right) \:.
\label{kbar3d2}
\end{equation}
Here, according to figure \ref{kernelaxi2}, the axial component changes sign for $t\approx0.7$, so that inertial effects decrease the axial drag at longer times. Comparing (\ref{kbar3d2}) and (\ref{kbar3d}) makes it clear that the steady-state value of each component is totally different in the two flows, which underlines the nonlinear nature of advective effects. }}
\bibliographystyle{jfm}
\bibliography{biblio}

\end{document}